\documentclass[12pt,a4paper]{article}
\usepackage{a4wide}
\usepackage{amsmath}
\usepackage{amssymb}
\usepackage{amsfonts}
\usepackage{epsfig}
\usepackage{subfigure}
\usepackage{exscale}
\usepackage{float}
\usepackage{bbm}
\usepackage[numbers,sort&compress]{natbib}

\newcommand{\Z}{{\mathbb{Z}}}
\newcommand{\R}{{\mathbb{R}}}

\newcommand{\1}{{\mathbbm{1}}}

\makeatletter
\@addtoreset{equation}{section}
\makeatother

\title{From Doubled Chern-Simons-Maxwell Lattice Gauge Theory to Extensions of
the Toric Code}

\author{T.\ Z.\ Olesen$^a$, N.\ D.\ Vlasii$^{a,b}$, and U.-J.\ Wiese$^a$ \\ \\
{\small $^a$ Albert Einstein Center for Fundamental Physics, 
Institute for Theoretical Physics} \\
{\small Bern University, Sidlerstrasse 5, CH-3012 Bern, Switzerland} \\
{\small $^b$ Bogolyubov Institute for Theoretical Physics, 
National Academy of Sciences of Ukraine} \\
{\small 14-b Metrologichna Str., Kyiv, 03680, Ukraine}}

\begin{document} 

\maketitle

\begin{abstract} \normalsize

We regularize compact and non-compact Abelian Chern-Si\-mons-Max\-well theories 
on a spatial lattice using the Hamiltonian formulation. We consider a doubled 
theory with gauge fields living on a lattice and its dual lattice. The 
Hilbert space of the theory is a product of local Hilbert spaces, each 
associated with a link and the corresponding dual link. The two electric field 
operators associated with the link-pair do not commute. In the 
non-compact case with gauge group $\R$, each local Hilbert space is analogous to
the one of a charged ``particle'' moving in the link-pair group space $\R^2$ in 
a constant ``magnetic'' background field. In the compact case, the link-pair 
group space is a torus $U(1)^2$ threaded by $k$ units of quantized ``magnetic'' 
flux, with $k$ being the level of the Chern-Simons 
theory. The holonomies of the torus $U(1)^2$ give rise to two self-adjoint 
extension parameters, which form two non-dynamical background lattice gauge 
fields that explicitly break the manifest gauge symmetry from $U(1)$ to $\Z(k)$.
The local Hilbert space of a link-pair then decomposes into representations of 
a magnetic translation group. In the pure Chern-Simons limit of a large 
``photon'' mass, this results in a $\Z(k)$-symmetric variant of Kitaev's toric 
code, self-adjointly extended by the two non-dynamical background lattice gauge 
fields. Electric charges on the original lattice and on the dual lattice obey 
mutually anyonic statistics with the statistics angle $\frac{2 \pi}{k}$. 
Non-Abelian $U(k)$ Berry gauge fields that arise from the self-adjoint extension
parameters may be interesting in the context of quantum information processing.

\end{abstract}

\newpage
 
\section{Introduction}

Gauge theories in two spatial dimensions may contain a Chern-Simons term 
\cite{Che74,Des82,Dun89,Jac90,Dun99} which explicitly breaks parity and 
time-reversal symmetry. In these theories the gauge field acquires a mass and 
the charged particles obey fractional anyonic statistics 
\cite{Wil83,Fro89,Wil90,Ler92}. Non-Abelian Chern-Simons theories have intricate
relations to knot theory, the Jones polynomials \cite{Jon87}, and to 
2-dimensional conformal field theories \cite{Wit88,Wit89}. Abelian Chern-Simons 
theories have been used to facilitate bosonization or Fermi-Bose transmutation 
in $(2+1)$ dimensions \cite{Cos89,Lue89,Eli92,Eli92a}. Furthermore, Chern-Simons
gauge theories are of central importance in the context of the fractional 
quantum Hall effect \cite{Zha89,Jai89,Zha92} and other condensed matter systems
\cite{Fen05,Kou08,Pal13}. They also play an important role for topological 
quantum computation \cite{Kit03,Den02,Fre03,Fre04,Nay08,Bre08}. By braiding 
world-lines of anyonic quasi-particles, one can accumulate appropriate 
non-Abelian Berry phases \cite{Ber84,Sim83}, which can encode quantum 
information. When the anyons obey a sufficiently complex version of non-Abelian 
braid statistics, they can be employed to realize the quantum gates that are 
sufficient to realize universal quantum computation \cite{Deu85,Cir95,Sho97}. 
The idea of topological quantum computation is attractive because the quantum 
information is then naturally protected from decoherence by the 
topological nature of the non-Abelian Berry phases. In particular, information 
is not stored locally but is distributed throughout the entire system. The toric
code is a $(2+1)$-d $\Z(2)$ lattice gauge theory which can be used as a 
topologically protected storage device for quantum information \cite{Kit03}. 
This theory has charges and dual charges with mutually anyonic statistics.

In this paper, we derive extensions of the toric code from a doubled compact 
lattice Chern-Simons-Maxwell theory with Abelian gauge group $U(1)$. Similar to 
fermions, topologically massive gauge fields also suffer from a lattice 
doubling problem. Here we are not trying to circumvent this problem but work
with a lattice gauge field and an independent gauge field associated with the
dual lattice. The fundamental gauge degrees of freedom are then associated with 
a cross formed by a link and its corresponding dual link. The Chern-Simons term
couples the two lattices and implies that the canonically conjugate momenta 
(i.e.\ the electric field strengths) of the original and the dual gauge field
do not commute. In ordinary lattice gauge theories 
\cite{Wil74,Kog75,Sus77,Kog83} the field algebra is link-based. This framework 
has also been used in studies of Chern-Simons gauge theories on the lattice 
\cite{Lue89,Mul90,Ada97,Ber00,Dou05a,Bai09}. In our lattice formulation of a 
doubled Chern-Simons gauge theories, on the other hand, the field algebra is 
cross-based. Such a system was first investigated in \cite{Kan91}. Here we 
concentrate on the relation of this theory to the toric code. 

In ordinary lattice gauge theory with a link-based field algebra every link has 
a ``mechanical'' analog. It behaves like a ``particle'' moving in the group 
space. For example, the dynamics of a link variable in a compact Abelian $U(1)$ 
lattice gauge theory is analogous to the one of a quantum mechanical particle 
moving on a circle. Similarly, in our unconventional lattice gauge theory with 
a cross-based field algebra, the ``mechanical'' analog of each cross is a 
charged ``particle'' moving on a 2-dimensional group space torus $U(1)^2$ 
threaded by an abstract ``magnetic'' field \cite{Zai89,Che96,AlH09}.
The Dirac quantization condition for the abstract ``magnetic'' flux then
implies the quantization of the level $k$ --- the prefactor of the Chern-Simons
term. Interestingly, the corresponding cross-based Hamiltonian has two
self-adjoint extension parameters, which naturally enter the quantum theory as
external parameters, while the classical theory is insensitive to these
parameters. Remarkably, the self-adjoint extension parameters (which are 
associated with the links and the dual links) themselves form two non-dynamical 
$U(1)$ lattice gauge fields. This reduces the manifest gauge symmetry of the 
quantum theory from $U(1)$ to $\Z(k)$. This quantum mechanical breaking of the 
gauge symmetry could be called an ``anomaly''. However, it is more like the 
explicit breaking of CP invariance caused by a non-zero $\theta$-vacuum angle 
in 4-dimensional non-Abelian gauge theories. It is intriguing that the 
quantized doubled Chern-Simons-Maxwell lattice gauge theory dynamically reduces 
its manifest gauge symmetry from $U(1)$ to the $\Z(k)$ gauge group of an
extended toric code. The full dynamics reduces to the one of the toric 
code in the pure Chern-Simons limit of infinite ``photon'' mass.

The main purpose of this paper is to perform a detailed mathematical derivation
of an extended version of the $\Z(k)$ variant of the toric code as a limit of 
doubled $U(1)$ Chern-Simons-Maxwell lattice gauge theory. We put particular 
emphasis on the role played by the self-adjoint extension parameters of the 
cross-based Hamiltonian. This embeds the toric code in a wider theoretical 
framework and thus provides a broader perspective on quantum information 
processing. In particular, it would be interesting to investigate whether 
Berry phases that arise from adiabatic changes of the external self-adjoint 
extension parameters can be utilized for quantum information processing. In this
paper, we do not yet address these questions. We also view the present paper as 
a first step towards similar studies in the context of non-Abelian Chern-Simons 
theories on the lattice. Variants of the toric code with a discrete non-Abelian 
gauge group \cite{Dou05}, which are discussed in the context of topological 
quantum computation, may be related to Chern-Simons gauge theories with 
continuous gauge groups in a similar manner.

The rest of the paper is organized as follows. In Section 2 we discuss 
Chern-Simons-Maxwell gauge theory of a single Abelian gauge field in the 
continuum, while in Section 3 we investigate the doubled theory with two
Abelian gauge fields. In Section 4 we regularize this theory on a lattice and
its dual lattice, with a cross-based field algebra using non-compact Abelian
lattice gauge fields. In particular, we discuss the mutual anyonic statistics 
of the charges moving on the original and the dual lattice. In Section 5 we turn
to compact Abelian gauge fields, which leads to the quantization of the level
$k$ as well as to the generation of a non-dynamical background lattice gauge 
field formed by the self-adjoint extension parameters of the cross-based
Hamiltonian. In the limit of a large ``photon'' mass this theory reduces to the 
toric code. Finally, Section 6 contains our conclusions.

\section{Chern-Simons-Maxwell Theory of a Single \\ Abelian Gauge Field in the
Continuum}

In this section we investigate Abelian Chern-Simons-Maxwell theory in the
continuum. After considering a single Abelian gauge field, in the next section
we will study a doubled theory which will turn out to arise in the continuum 
limit of the lattice theory that we discuss later.

\subsection{Lagrangian and Hamiltonian}

Let us consider a $(2+1)$-d Abelian gauge field $A_\mu(x)$ with the 
Chern-Simons-Maxwell Lagrangian
\begin{equation}
{\cal L} = - \frac{1}{4 e^2} F_{\mu\nu} F^{\mu\nu} +
\frac{k}{4 \pi} \epsilon_{\mu\nu\rho} A^\mu \partial^\nu A^\rho,
\end{equation}
where $e$ is the electric charge and the metric is 
$g_{\mu\nu} = \mbox{diag}(1,-1,-1)$. The field strength is given by
$F_{\mu\nu}(x) = \partial_\mu A_\nu(x) - \partial_\nu A_\mu(x)$. Without loss of 
generality, in the following we assume that the prefactor of the Chern-Simons
density $k > 0$. It is interesting to note that the Chern-Simons density is not 
gauge invariant. However, its variation under a gauge transformation 
\begin{equation}
A'_\mu(x) = A_\mu(x) - \partial_\mu \chi(x),
\end{equation}
is a total divergence
\begin{equation}
\delta {\cal L} = {\cal L}' - {\cal L} = - \partial^\mu 
\left(\frac{k}{4 \pi} \epsilon_{\mu\nu\rho} \chi \partial^\nu A^\rho\right).
\end{equation}
It should be noted that the Chern-Simons term explicitly breaks both 
time-reversal and parity (i.e.\ the reflection on a spatial axis). In order to 
derive the corresponding Hamiltonian, we now fix to the temporal gauge 
$A_0(x) = 0$, which implies
\begin{equation}
{\cal L} = \frac{1}{2 e^2} \dot A_i^2 - 
\frac{1}{2 e^2}(\epsilon_{ij} \partial_i A_j)^2 + 
\frac{k}{4 \pi} \epsilon_{ij} \dot A_i A_j.
\end{equation}
The momentum canonically conjugate to $A_i$ is then given by
\begin{equation}
\Pi_i(x) = \frac{\delta {\cal L}}{\delta \dot A_i(x)} = 
\frac{1}{e^2} \dot A_i(x) + \frac{k}{4 \pi} \epsilon_{ij} A_j(x),
\end{equation}
such that the classical Hamilton density takes the form
\begin{eqnarray}
{\cal H}&=&\Pi_i \dot A_i - {\cal L} = \frac{1}{2 e^2} \dot A_i^2 +
\frac{1}{2 e^2}(\epsilon_{ij} \partial_i A_j)^2 \nonumber \\
&=&\frac{e^2}{2} \left(\Pi_i - \frac{k}{4 \pi} \epsilon_{ij} A_j\right)^2 +
\frac{1}{2 e^2}(\epsilon_{ij} \partial_i A_j)^2 = 
\frac{e^2}{2} E_i^2 + \frac{1}{2 e^2} B^2.
\end{eqnarray}
Here we have identified the electric and magnetic fields as
\begin{equation}
E_i(x) = \Pi_i(x) - \frac{k}{4 \pi} \epsilon_{ij} A_j(x), \quad 
B(x) = \partial_1 A_2(x) - \partial_2 A_1(x).
\end{equation}

\subsection{Solutions of the Classical Equations of Motion}

Before we quantize the theory, we consider its classical equations of 
motion
\begin{equation}
\partial^\mu F_{\mu\nu}(x) + 
\frac{k e^2}{4 \pi} \epsilon_{\nu\rho\sigma} F^{\rho\sigma}(x) = 0,
\end{equation}
which in components take the form
\begin{equation}
\dot E_i(x) + \frac{1}{e^2} \epsilon_{ij} \partial_j B(x) + 
\frac{k e^2}{2 \pi} \epsilon_{ij} E_j(x) = 0, \quad
\partial_i E_i(x) + \frac{k}{2 \pi} B(x) = 0.
\end{equation}
The second equation is the Gauss law. In particular, the magnetic field acts 
like a charge density. In addition, the field strength obeys the Bianchi 
identity
\begin{equation}
\epsilon_{\mu\nu\rho} \partial^\mu F^{\nu\rho}(x) = 0.
\end{equation}

Let us make the simple plane wave ansatz
\begin{equation}
E_i(x) = C_i \cos(\vec p \cdot \vec x - \omega t) +
D_i \sin(\vec p \cdot \vec x - \omega t).
\end{equation}
Inserting this in the equations of motion, one obtains
\begin{equation}
\omega = \sqrt{M^2 + p^2}, \quad M = \frac{k e^2}{2 \pi}, \quad
C_i = c \epsilon_{ij} p_j - d \frac{M}{\omega} p_i, \quad
D_i = d \epsilon_{ij} p_j + c \frac{M}{\omega} p_i,
\end{equation}
where $c, d \in \R$ are arbitrary constants. We have identified $M$ as the 
topologically generated ``photon'' mass of the Abelian gauge field, which is 
proportional to the prefactor $k$ of the Chern-Simons term. In the absence of 
the Maxwell term, i.e.\ when $e^2 \rightarrow \infty$, the gauge field becomes 
infinitely heavy. In the pure Maxwell theory (with $k = 0$), on the other hand, 
the gauge field remains massless, and plane waves are purely transverse.

In order to quantize the theory, we now impose canonical commutation relations
\begin{equation}
[\Pi_i(x),A_j(y)] = - i \delta_{ij} \delta(x - y),
\end{equation}
which imply the following commutation relations between the electric and 
magnetic fields
\begin{eqnarray}
&&[E_i(x),E_j(y)] = - i \frac{k}{2 \pi} \epsilon_{ij} \delta(x - y), \nonumber \\
&&[B(x),B(y)] = 0, \nonumber \\
&&[E_i(x),B(y)] = i \epsilon_{ij} \partial_j \delta(x - y).
\end{eqnarray}
The derivative in the third equation is with respect to $x$ (not $y$). As a 
consequence of the Chern-Simons term, the two components of the electric field 
do not commute with each other. It is easy to check that the Hamiltonian 
\begin{equation}
H = \int d^2x \left(\frac{e^2}{2} E_i^2 + \frac{1}{2 e^2} B^2\right)
\end{equation}
commutes with the infinitesimal generators of local gauge transformations
\begin{equation}
G(x) = \partial_i E_i(x) + \frac{k}{2 \pi} B(x), \quad [G(x),H] = 0.
\end{equation}
As usual in a gauge theory, physical states $|\Psi\rangle $ must obey the Gauss 
law $G(x)|\Psi\rangle = 0$. Upon quantization, the classical plane wave 
solutions then turn into free ``photon'' states of mass 
$M = \frac{k e^2}{2 \pi}$.

\section{Doubled Continuum Theory with two Abelian Gauge Fields}

In the next section, we will regularize Chern-Simons-Maxwell theory on a 
particular lattice, which will result in a doubling problem, similar to the 
well-known lattice fermion doubling problem. The continuum limit of that lattice
theory is a doubled Chern-Simons-Maxwell theory with two Abelian gauge fields,
which we first investigate directly in the continuum.

\subsection{Lagrangian and Hamiltonian}

We now consider two Abelian gauge fields $A_\mu(x)$ and $\widetilde A_\mu(x)$
with the Lagrangian
\begin{equation}
{\cal L} = - \frac{1}{4 e^2} F_{\mu\nu} F^{\mu\nu} 
- \frac{1}{4 \widetilde e^2} \widetilde F_{\mu\nu} \widetilde F^{\mu\nu} +
\frac{k}{4 \pi} \epsilon_{\mu\nu\rho} (A^\mu \partial^\nu \widetilde A^\rho +
\widetilde A^\mu \partial^\nu A^\rho).
\end{equation}
Without loss of generality one can assume that the dual charge equals the
original charge (i.e.\ $\widetilde e = e$). If this is not the case a priori,
one can achieve this by a simple rescaling of the fields. Unlike the theory
with a single gauge field, the doubled theory is invariant under time-reversal
and parity, because one can treat $\widetilde A_\mu(x)$ as a pseudo-vector.

Again, fixing to the temporal gauge $A_0(x) = 0$, $\widetilde A_0(x) = 0$, one 
obtains
\begin{equation}
{\cal L} = \frac{1}{2 e^2} \dot A_i^2 - 
\frac{1}{2 e^2}(\epsilon_{ij} \partial_i A_j)^2 +
\frac{1}{2 e^2} \dot {\widetilde A}_i^2 - 
\frac{1}{2 e^2}(\epsilon_{ij} \partial_i \widetilde A_j)^2 +
\frac{k}{4 \pi} \epsilon_{ij} 
(\dot A_i \widetilde A_j + \dot {\widetilde A}_i A_j),
\end{equation}
which yields the following canonically conjugate momenta
\begin{eqnarray}
&&\Pi_i(x) = \frac{\delta {\cal L}}{\delta \dot A_i(x)} = 
\frac{1}{e^2} \dot A_i(x) + \frac{k}{4 \pi} \epsilon_{ij} \widetilde A_j(x),
\nonumber \\
&&\widetilde \Pi_i(x) = \frac{\delta {\cal L}}{\delta \dot {\widetilde A}_i(x)} 
= \frac{1}{e^2} \dot {\widetilde A}_i(x) + \frac{k}{4 \pi} \epsilon_{ij} A_j(x).
\end{eqnarray}
The classical Hamilton density then takes the form
\begin{eqnarray}
{\cal H}&=&\Pi_i \dot A_i + \widetilde \Pi_i \dot {\widetilde A}_i - {\cal L} = 
\frac{1}{2 e^2} \dot A_i^2 +
\frac{1}{2 e^2}(\epsilon_{ij} \partial_i A_j)^2 +
\frac{1}{2 e^2} \dot {\widetilde A}_i^2 +
\frac{1}{2 e^2}(\epsilon_{ij} \partial_i \widetilde A_j)^2 \nonumber \\
&=&
\frac{e^2}{2} \left(\Pi_i - \frac{k}{4 \pi} \epsilon_{ij} \widetilde A_j\right)^2
+ \frac{1}{2 e^2}(\epsilon_{ij} \partial_i A_j)^2 +
\frac{e^2}{2} \left(\widetilde \Pi_i - \frac{k}{4 \pi} \epsilon_{ij} A_j\right)^2
+ \frac{1}{2 e^2}(\epsilon_{ij} \partial_i \widetilde A_j)^2 \nonumber \\
&=&\frac{e^2}{2} E_i^2 + \frac{1}{2 e^2} B^2 +
\frac{e^2}{2} \widetilde E_i^2 + \frac{1}{2 e^2} \widetilde B^2.
\end{eqnarray}
In this case, the electric and magnetic fields are given by
\begin{eqnarray}
&&E_i(x) = \Pi_i(x) - \frac{k}{4 \pi} \epsilon_{ij} \widetilde A_j(x), \quad 
B(x) = \partial_1 A_2(x) - \partial_2 A_1(x), \nonumber \\
&&\widetilde E_i(x) = \widetilde \Pi_i(x) - \frac{k}{4 \pi} \epsilon_{ij} A_j(x),
\quad 
\widetilde B(x) = \partial_1 \widetilde A_2(x) - \partial_2 \widetilde A_1(x).
\end{eqnarray}

\subsection{Solutions of the Classical Equations of Motion}

The classical equations of motion of the doubled theory are
\begin{equation}
\partial^\mu F_{\mu\nu}(x) + 
\frac{k e^2}{4 \pi} \epsilon_{\nu\rho\sigma} \widetilde F^{\rho\sigma}(x) = 0, \quad
\partial^\mu \widetilde F_{\mu\nu}(x) + 
\frac{k e^2}{4 \pi} \epsilon_{\nu\rho\sigma} F^{\rho\sigma}(x) = 0,
\end{equation}
which in components take the form
\begin{eqnarray}
&&\dot E_i(x) + \frac{1}{e^2} \epsilon_{ij} \partial_j B(x) + 
\frac{k e^2}{2 \pi} \epsilon_{ij} \widetilde E_j(x) = 0, \quad
\partial_i E_i(x) + \frac{k}{2 \pi} \widetilde B(x) = 0, \nonumber \\
&&\dot {\widetilde E}_i(x) + 
\frac{1}{e^2} \epsilon_{ij} \partial_j \widetilde B(x) + 
\frac{k e^2}{2 \pi} \epsilon_{ij} E_j(x) = 0, \quad
\partial_i \widetilde E_i(x) + \frac{k}{2 \pi} B(x) = 0.
\end{eqnarray}
As before, we make the plane wave ansatz
\begin{eqnarray}
E_i(x) = C_i \cos(\vec p \cdot \vec x - \omega t) +
D_i \sin(\vec p \cdot \vec x - \omega t), \nonumber \\
\widetilde E_i(x) = \widetilde C_i \cos(\vec p \cdot \vec x - \omega t) +
\widetilde D_i \sin(\vec p \cdot \vec x - \omega t),
\end{eqnarray}
which again implies $\omega = \sqrt{M^2 + p^2}$, and $M = \frac{k e^2}{2 \pi}$,
as well as
\begin{eqnarray}
&&C_i = c \epsilon_{ij} p_j - \widetilde d \frac{M}{\omega} p_i, \quad
D_i = d \epsilon_{ij} p_j + \widetilde c \frac{M}{\omega} p_i, \nonumber \\
&&\widetilde C_i = \widetilde c \epsilon_{ij} p_j - d \frac{M}{\omega} p_i, \quad
\widetilde D_i = \widetilde d \epsilon_{ij} p_j + c \frac{M}{\omega} p_i.
\end{eqnarray}
In this case, there are four independent amplitudes 
$c, d, \widetilde c, \widetilde d \in \R$, and thus there are two independent
``photon'' modes, both with the same mass $M$.

\subsection{Canonical Quantization}

Canonical quantization now amounts to
\begin{eqnarray}
&&[\Pi_i(x),A_j(y)] = [\widetilde \Pi_i(x),\widetilde A_j(y)] = 
- i \delta_{ij} \delta(x - y), 
\nonumber \\
&&[\Pi_i(x),\widetilde A_j(y)] = [\widetilde \Pi_i(x),A_j(y)] = 0,
\end{eqnarray}
which imply the following commutation relations between the electric and 
magnetic fields
\begin{eqnarray}
&&[E_i(x),\widetilde E_j(y)] = [\widetilde E_i(x),E_j(y)] = 
- i \frac{k}{2 \pi} \epsilon_{ij} \delta(x - y), \nonumber \\
&&[E_i(x),E_j(y)] = [\widetilde E_i(x),\widetilde E_j(y)] = 0, \nonumber \\
&&[B(x),B(y)] = [\widetilde B(x),\widetilde B(y)] = [B(x),\widetilde B(y)] = 0, 
\nonumber \\
&&[E_i(x),B(y)] = [\widetilde E_i(x),\widetilde B(y)] = 
i \epsilon_{ij} \partial_j \delta(x - y), \nonumber \\
&&[E_i(x),\widetilde B(y)] = [\widetilde E_i(x),B(y)] = 0.
\end{eqnarray}
The Hamiltonian now takes the form
\begin{equation}
H = \int d^2x \left(\frac{e^2}{2} E_i^2 + \frac{1}{2 e^2} B^2 +
\frac{e^2}{2} \widetilde E_i^2 + \frac{1}{2 e^2} \widetilde B^2\right).
\end{equation}
It commutes with the infinitesimal generators of two local gauge transformations
\begin{equation}
G(x) = \partial_i E_i(x) + \frac{k}{2 \pi} \widetilde B(x), \quad
\widetilde G(x) = \partial_i \widetilde E_i(x) + \frac{k}{2 \pi} B(x), \quad 
[G(x),H] = [\widetilde G(x),H] = 0.
\end{equation}
The Gauss law now constrains physical states to
$G(x)|\Psi\rangle = \widetilde G(x)|\Psi\rangle = 0$. The classical plane wave 
solutions then turn into the states of two free ``photons'', both of mass 
$M = \frac{k e^2}{2 \pi}$.

\section{Non-compact Chern-Simons-Maxwell Theory \\ on the Lattice}
 
In this section, we regularize Chern-Simons-Maxwell theory on a particular
spatial lattice. We work in the Hamiltonian formulation and thus leave time 
continuous. As we will see, the lattice theory suffers from a doubling problem 
and thus reduces to the doubled Chern-Simons-Maxwell theory in the continuum 
limit.

\subsection{Cross-based Degrees of Freedom}

For simplicity, we consider a square lattice. The generalization to other 
lattice geometries is straightforward but not very illuminating. Most previous 
lattice constructions of Chern-Simons theories have placed all dynamical gauge 
degrees of freedom on the same lattice. Here, similar to the lattice geometry 
used in \cite{Kan91}, we simultaneously place independent gauge degrees of 
freedom on the original as well as on the dual lattice. This is natural from a 
geometric point of view, but it inevitably leads to the doubling of the massive 
``photon'' mode. This is no problem in the present context, because it is our 
goal to explicitly connect lattice Chern-Simons-Maxwell theory to the toric 
code, which indeed requires to start from the doubled theory.
\begin{figure}[tbp]
\begin{center}
\includegraphics[width=0.5\textwidth]{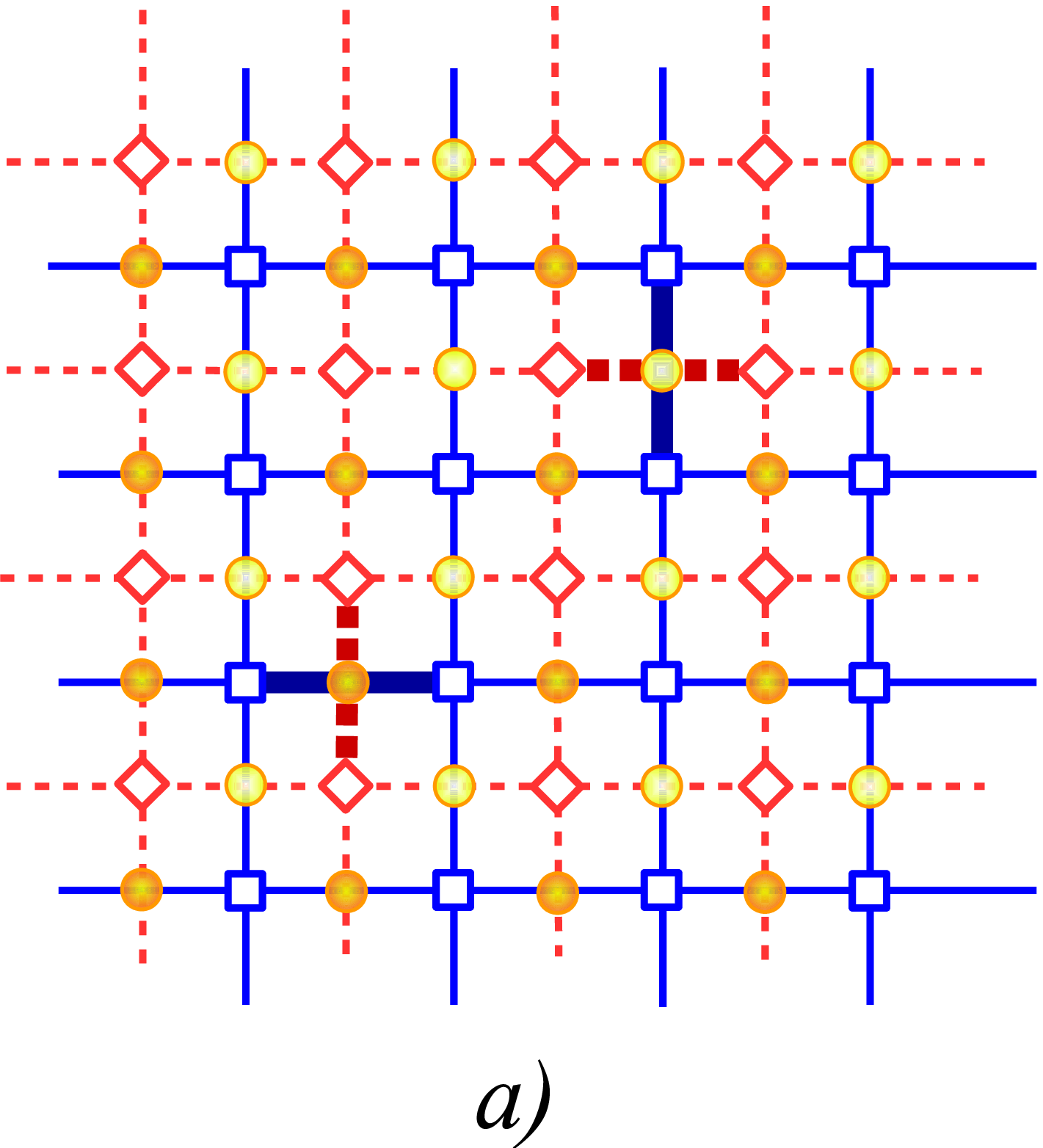} \hspace{1cm}
\includegraphics[width=0.21\textwidth]{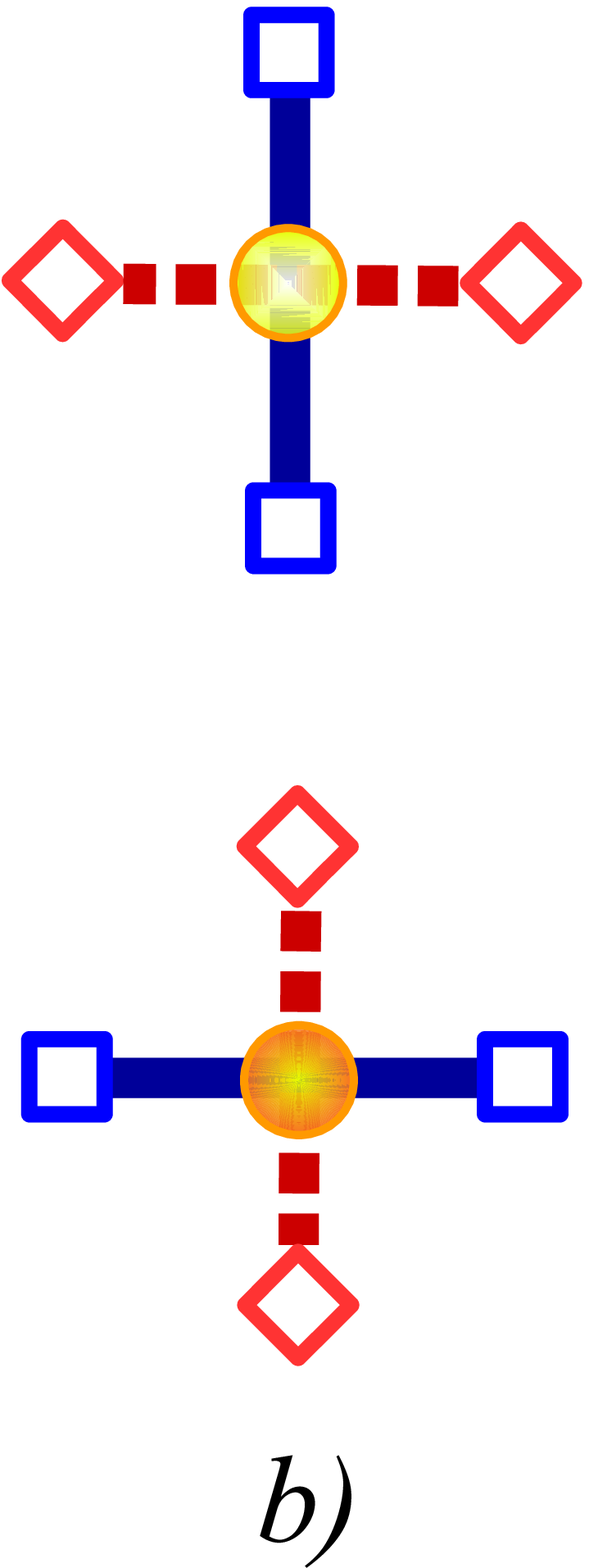}
\end{center}
\caption{[Color online] \textit{a) Square lattice $\Lambda$ (squares, solid
links) together with its dual lattice $\widetilde \Lambda$ (diamonds, dashed
links). The gauge fields $A_{x,i}$ and $\widetilde A_{x,i}$ are associated with
two links forming a cross centered at the point $x$. The cross centers $x$ do
not belong to the lattices $\Lambda$ or $\widetilde \Lambda$, but form two
other sublattices $X$ (filled circles) and $\widetilde X$ (open circles). b) 
Two types of crosses of link-pairs, $A_{x,1}, \widetilde A_{x,2}$ and 
$\widetilde A_{x,1}, A_{x,2}$, whose cross centers $x$ belong to the two distinct 
sublattices $X$ and $\widetilde X$, respectively.}}
\label{lattice}
\end{figure}

In standard lattice gauge theory, the basic gauge field algebra is link-based, 
i.e.\ the gauge field variables as well as their canonically conjugate electric 
field operators are defined on a link. In particular, field operators residing 
on different links commute with one another. As we have seen, in the doubled
Chern-Simons-Maxwell theory the canonically conjugate variable to $A_i$ is 
$\Pi_i = \frac{1}{e^2} \dot A_i + \frac{k}{4 \pi} \epsilon_{ij} \widetilde A_j$. 
For the corresponding lattice theory, this implies that the gauge variables 
associated with different links no longer necessarily commute. As we will see, 
it is natural to associate the non-commuting gauge field variables with a link 
and the corresponding dual link. Consequently, the basic gauge algebra of 
doubled lattice Chern-Simons theory is not based on a single link, but on a 
pair of links which are dual to each other. The original square lattice 
together with its dual lattice is illustrated in Fig.\ref{lattice}.

Let us consider a link and the corresponding dual link which form a cross
centered at a point $x$. Note that $x$ is neither a site of the original nor 
of the dual lattice, but just an intersection point of the link and its dual 
link on a sublattice $X$ or $\widetilde X$ (cf.\ Fig.\ref{lattice}). We denote 
the non-compact vector potential on the original link by $A_{x,i}$. The vector 
potential on the corresponding dual link is then given by 
$\epsilon_{ij} \widetilde A_{x,j}$. It should be noted that for a given cross,
the direction $i$ of the original link variable $A_{x,i}$ and the orthogonal 
direction $j$ of the dual link variable $\widetilde A_{x,j}$ are determined by
the location of the intersection point $x$ on sublattice $X$ or $\widetilde X$.

\subsection{Lagrangian and Hamiltonian}

In the continuum the integrated Lagrange density (i.e.\ the Lagrange function)
is given by
\begin{equation}
L = \int d^2x \ \left[\frac{1}{2 e^2} \dot A_i^2 - \frac{1}{2 e^2} B^2 +
\frac{1}{2 e^2} \dot {\widetilde A}_i^2 - \frac{1}{2 e^2} \widetilde B^2 +
\frac{k}{4 \pi} \epsilon_{ij} 
(\dot A_i \widetilde A_j + \dot {\widetilde A}_i A_j)\right].
\end{equation}
Correspondingly, on the lattice we obtain
\begin{eqnarray}
L&=&\sum_{x \in X} a^2 \left[\frac{1}{2 e^2} \dot A_{x,1}^2 +
\frac{1}{2 e^2} \dot {\widetilde A}_{x,2}^2 + 
\frac{k}{4 \pi}(\dot A_{x,1} \widetilde A_{x,2} - 
\dot {\widetilde A}_{x,2} A_{x,1})\right]
\nonumber \\
&+&\sum_{x \in \widetilde X} a^2 \left[\frac{1}{2 e^2} \dot A_{x,2}^2 +
\frac{1}{2 e^2} \dot {\widetilde A}_{x,1}^2 + 
\frac{k}{4 \pi}(\dot {\widetilde A}_{x,1} A_{x,2} - 
\dot A_{x,2} \widetilde A_{x,1})\right]
\nonumber \\
&-&\sum_{x \in \widetilde \Lambda} a^2 \frac{1}{2 e^2} B_x^2 -
\sum_{x \in \Lambda} a^2 \frac{1}{2 e^2} \widetilde B_x^2.
\end{eqnarray}
In the sums over the sublattices $X$, $\widetilde X$, $\Lambda$, 
$\widetilde \Lambda$ the factors $a^2$ (where $a$ is the spacing of the 
original or the dual lattice) arises as the area of an elementary plaquette. On 
the lattice, the magnetic fields are defined as
\begin{eqnarray}
&&B_x = \frac{1}{a} \left(A_{x-\frac{a}{2}\hat 2,1} + A_{x+\frac{a}{2}\hat 1,2}
- A_{x+\frac{a}{2}\hat 2,1} - A_{x-\frac{a}{2}\hat 1,2}\right), \nonumber \\
&&\widetilde B_x = \frac{1}{a} \left(\widetilde A_{x-\frac{a}{2}\hat 2,1} + 
\widetilde A_{x+\frac{a}{2}\hat 1,2} - \widetilde A_{x+\frac{a}{2}\hat 2,1} - 
\widetilde A_{x-\frac{a}{2}\hat 1,2}\right),
\end{eqnarray}
with $\hat i$ being the unit-vector in the $i$-direction. Note that the magnetic
field $B_x$, which is built around a plaquette of the original lattice, is
indexed by the dual site $x \in \widetilde \Lambda$ at the center of this 
plaquette. The magnetic fields are invariant against lattice gauge 
transformations
\begin{equation}
A'_{x,i} = A_{x,i} - \frac{1}{a}
\left(\chi_{x+\frac{a}{2}\hat i} - \chi_{x-\frac{a}{2}\hat i}\right), \quad
\widetilde A'_{x,i} = \widetilde A_{x,i} - \frac{1}{a}
\left(\widetilde \chi_{x+\frac{a}{2}\hat i} - \widetilde \chi_{x-\frac{a}{2}\hat i}
\right).
\end{equation}

On the lattice, the canonically conjugate momenta are given by 
\begin{eqnarray}
&&\Pi_{x,i} = \frac{\partial L}{\partial \dot A_{x,i}} = 
a^2 \left(\frac{1}{e^2} \dot A_{x,i} + 
\frac{k}{4 \pi} \epsilon_{ij} \widetilde A_{x,j}\right), \nonumber \\
&&\widetilde \Pi_{x,i} = 
\frac{\partial L}{\partial \dot {\widetilde A}_{x,i}} 
= a^2 \left(\frac{1}{e^2} \dot {\widetilde A}_{x,i} + 
\frac{k}{4 \pi} \epsilon_{ij} A_{x,j}\right).
\end{eqnarray}
Note that on the lattice $\Pi_{x,i}$ has a different dimension than $\Pi_i(x)$ in
the continuum. The corresponding classical Hamilton function then takes the form
\begin{eqnarray}
\label{latticeH}
H&=&\sum_{x \in X} (\Pi_{x,1} \dot A_{x,1} + 
\widetilde \Pi_{x,2} \dot {\widetilde A}_{x,2}) +
\sum_{x \in \widetilde X} (\Pi_{x,2} \dot A_{x,2} + 
\widetilde \Pi_{x,1} \dot {\widetilde A}_{x,1}) - L \nonumber \\
&=&\frac{e^2}{2} \sum_{x \in X} a^2 (E_{x,1}^2 + \widetilde E_{x,2}^2) +
\frac{e^2}{2} \sum_{x \in \widetilde X} a^2
(E_{x,2}^2 + \widetilde E_{x,1}^2) \nonumber \\
&+&\frac{1}{2 e^2} \sum_{x \in \widetilde \Lambda} a^2 B_x^2 +
\frac{1}{2 e^2} \sum_{x \in \Lambda} a^2 \widetilde B_x^2.
\end{eqnarray}
In this case, the electric fields are given by
\begin{equation}
E_{x,i} = \frac{1}{a^2} \Pi_{x,i} - 
\frac{k}{4 \pi} \epsilon_{ij} \widetilde A_{x,j}, \quad
\widetilde E_{x,i} = \frac{1}{a^2} \widetilde \Pi_{x,i} - 
\frac{k}{4 \pi} \epsilon_{ij} A_{x,j}.
\end{equation}

\subsection{Solutions of the Classical Equations of Motion}

\begin{figure}[tbp]
\begin{center}
\includegraphics[width=0.3\textwidth]{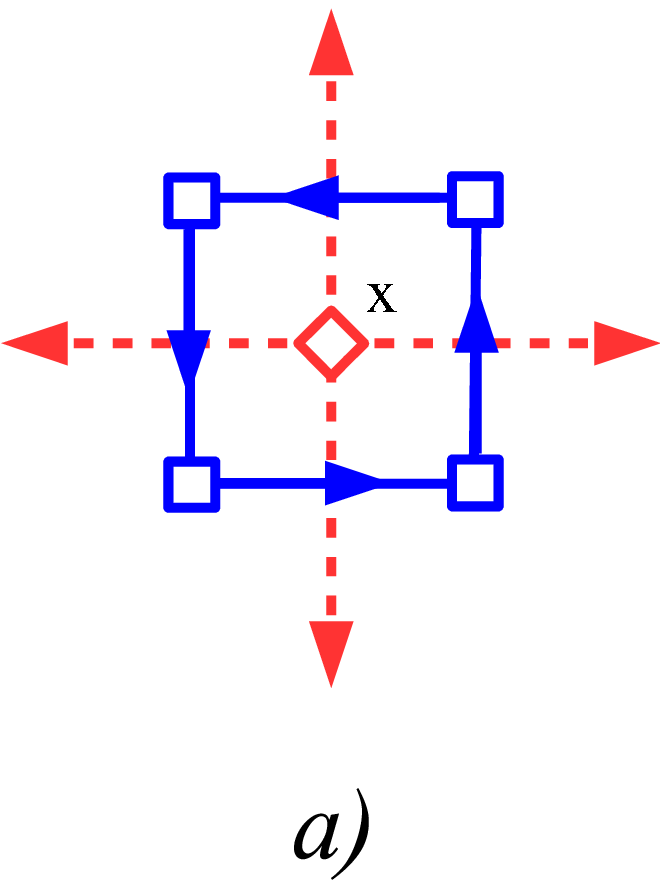} \hspace{0.4cm}
\includegraphics[width=0.5\textwidth]{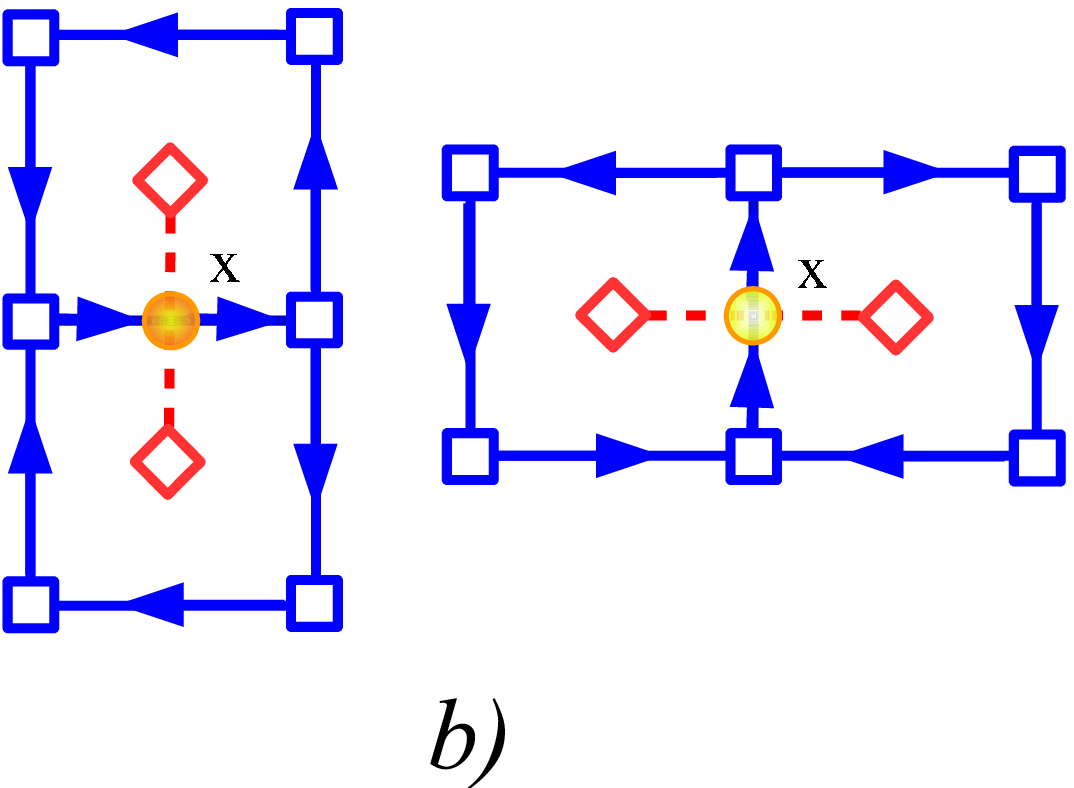}
\end{center}
\caption{[Color online] \textit{Geometry of the lattice equations of motion:
a) The Gauss law arises from the divergence of the electric field, i.e.\ from
the electric fluxes flowing out of a lattice point $x$, and from the magnetic 
field associated with a plaquette formed with the dual links. b) The other
two equations of motion relate the time-derivative of an electric field 
$\dot E_{x,i}$ on a link centered at $x$ to the dual electric field 
$\epsilon_{ij} \widetilde E_{x,j}$ as well as to the difference of magnetic 
fields residing on two nearest-neighbor plaquettes.}}
\label{Gauss}
\end{figure}

Before we quantize the theory, let us first consider the classical equations of
motion on the lattice, which in components take the form
\begin{eqnarray}
&&\sum_i \frac{1}{a} (E_{x+\frac{a}{2} \hat i,i} - E_{x-\frac{a}{2} \hat i,i})
+ \frac{k}{2 \pi} \widetilde B_x = 0, \nonumber \\
&&\dot E_{x,i} + 
\frac{1}{e^2} \epsilon_{ij} \frac{1}{a}(B_{x+\frac{a}{2} \hat j} - 
B_{x-\frac{a}{2} \hat j}) + 
\frac{k e^2}{2 \pi} \epsilon_{ij} \widetilde E_{x,j} = 0, \nonumber \\
&&\sum_i \frac{1}{a} 
(\widetilde E_{x+\frac{a}{2} \hat i,i} - \widetilde E_{x-\frac{a}{2} \hat i,i})
+ \frac{k}{2 \pi} B_x = 0, \nonumber \\
&&\dot {\widetilde E}_{x,i} + \frac{1}{e^2} \epsilon_{ij} 
\frac{1}{a}(\widetilde B_{x+\frac{a}{2} \hat j} - \widetilde B_{x-\frac{a}{2} \hat j}) +
\frac{k e^2}{2 \pi} \epsilon_{ij} E_{x,j} = 0.
\end{eqnarray}
The first equation corresponds to the lattice Gauss law, which is illustrated in
Fig.\ref{Gauss}a. The geometry underlying the second equation is depicted in
Fig.\ref{Gauss}b. 
As in the continuum, we make the plane wave ansatz
\begin{eqnarray}
E_{x,i} = C_i \cos(\vec p \cdot \vec x - \omega t) +
D_i \sin(\vec p \cdot \vec x - \omega t), \nonumber \\
\widetilde E_{x,i} = \widetilde C_i \cos(\vec p \cdot \vec x - \omega t) +
\widetilde D_i \sin(\vec p \cdot \vec x - \omega t).
\end{eqnarray}
Inserting this in the lattice equations of motion, we obtain the dispersion
relation
\begin{equation}
\omega = \sqrt{M^2 + \hat p^2}, \quad M = \frac{k e^2}{2 \pi}, \quad
\hat p_i = \frac{2}{a} \sin\frac{p_i a}{2}.
\end{equation}
Note that the mass of the gauge field is the same as in the continuum, while
the energy-momentum dispersion relation is distorted by lattice artifacts for
lattice momenta $p$ close to the edge of the Brillouin zone. In analogy to the
continuum result, we obtain
\begin{eqnarray}
&&C_i = c \epsilon_{ij} \hat p_j - \widetilde d \frac{M}{\omega} \hat p_i, \quad
D_i = d \epsilon_{ij} \hat p_j + \widetilde c \frac{M}{\omega} \hat p_i, 
\nonumber \\
&&\widetilde C_i = \widetilde c \epsilon_{ij} \hat p_j - 
d \frac{M}{\omega} \hat p_i, \quad
\widetilde D_i = \widetilde d \epsilon_{ij} \hat p_j + 
c \frac{M}{\omega} \hat p_i.
\end{eqnarray}
As in the continuum, there are four independent amplitudes 
$c, d, \widetilde c, \widetilde d \in \R$, and thus the massive ``photon'' mode 
is again doubled.

\subsection{Canonical Quantization}

Upon canonical quantization we now obtain
\begin{eqnarray}
&&[\Pi_{x,i},A_{y,j}] = [\widetilde \Pi_{x,i},\widetilde A_{y,j}] = 
- i \delta_{ij} \delta_{xy}, 
\nonumber \\
&&[\Pi_{x,i},\widetilde A_{y,j}] = [\widetilde \Pi_{x,i},A_{y,j}] = 0.
\end{eqnarray}
The resulting commutation relations for the vector potential and the electric 
field are
\begin{eqnarray}
\label{crosscom}
&&[A_{x,i},A_{y,j}] = [\widetilde A_{x,i},\widetilde A_{y,j}] = 
[A_{x,i},\widetilde A_{y,j}] = 0, \nonumber \\
&&[E_{x,i},A_{y,j}] = [\widetilde E_{x,i},\widetilde A_{y,j}] = 
- i \delta_{ij} \frac{1}{a^2} \delta_{xy}, \nonumber \\
&&[E_{x,i},\widetilde A_{y,j}] = [\widetilde E_{x,i},A_{y,j}] = 0, \nonumber \\
&&[E_{x,i},\widetilde E_{y,j}] = [\widetilde E_{x,i},E_{y,j}] = 
- i \frac{k}{2 \pi} \epsilon_{ij} \frac{1}{a^2} \delta_{xy}, \nonumber \\
&&[E_{x,i},E_{y,j}] = [\widetilde E_{x,i},\widetilde E_{y,j}] = 0.
\end{eqnarray}
In the continuum limit $a \rightarrow 0$, $\frac{1}{a^2} \delta_{xy}$ approaches 
$\delta(x-y)$. It is interesting to note that the commutation relations between
$A_{x,i}$ and $\epsilon_{ij} \widetilde A_{x,j}$ with $E_{x,i}$ and 
$\epsilon_{ij} \widetilde E_{x,j}$, which reside on the cross formed by a link 
and its dual link, correspond to those of a charged particle moving in a 2-d
plane in a constant magnetic field $\frac{k}{2 \pi}$. In this ``mechanical''
analog, $A_{x,i}$ and $\epsilon_{ij} \widetilde A_{x,j}$ play the role of the
particle's $x$- and $y$-coordinate, while $E_{x,i}$ and 
$\epsilon_{ij} \widetilde E_{x,j}$ represent the corresponding momenta. Like for
a charged particle in a magnetic field, the commutator of the momenta is 
non-zero and given by the abstract ``magnetic'' field $\frac{k}{2 \pi}$. In the
next section, we will compactify the gauge field from $\R$ to $U(1)$, which
will naturally lead to a quantization of the abstract ``magnetic'' field in 
integer units $k \in \Z$. For the moment, however, $k$ is not quantized.

\begin{figure}[tbp]
\begin{center}
\includegraphics[width=0.35\textwidth]{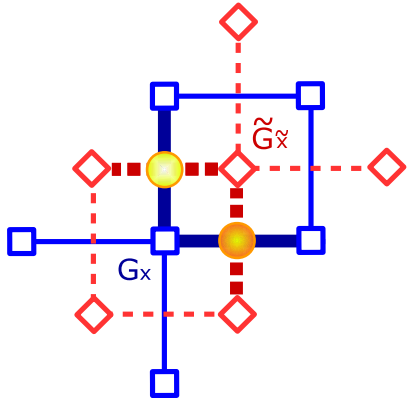}
\end{center}
\caption{[Color online] \textit{Consistency of the Gauss laws on the original
and on the dual lattice. The gauge generators $G_x$ and 
$\widetilde G_{\widetilde x}$ (with 
$\widetilde x = x + \frac{a}{2} \hat 1 + \frac{a}{2} \hat 2$) commute, i.e.\
$[G_x,\widetilde G_{\widetilde x}] = 0$. This is non-trivial, because, e.g., the 
electric fields $E_{x+\frac{a}{2} \hat 1,1}$ and 
$\widetilde E_{\widetilde x - \frac{a}{2} \hat 2,2}$ that contribute to $G_x$ and 
$\widetilde G_{\widetilde x}$ do not commute since they belong to the same cross 
formed by a link and its dual link.}}
\label{Gaussconsistency}
\end{figure}

The commutation relations between electric and magnetic fields
now take the form
\begin{eqnarray}
&&[E_{x,i},B_y] = [\widetilde E_{x,i},\widetilde B_y] = i \epsilon_{ij} 
\frac{1}{a^3} (\delta_{x,y+\frac{a}{2} \hat j} - \delta_{x,y-\frac{a}{2} \hat j}), 
\nonumber \\
&&[E_{x,i},\widetilde B_y] = [\widetilde E_{x,i},B_y] = 0, \nonumber \\
&&[B_x,B_y] = [\widetilde B_x,\widetilde B_y] = [B_x,\widetilde B_y] = 0.
\end{eqnarray}
The Hamiltonian of eq.(\ref{latticeH}) commutes with the infinitesimal 
generators of local gauge transformations, both on the original and on the dual
lattice
\begin{eqnarray}
\label{gauge}
&&\frac{G_x}{a^2} = 
\sum_i \frac{1}{a} (E_{x+\frac{a}{2} \hat i,i} - E_{x-\frac{a}{2} \hat i,i}) + 
\frac{k}{2 \pi} \widetilde B_x, \quad [G_x,H] = 0, \quad x \in \Lambda,
\nonumber \\
&&\frac{\widetilde G_x}{a^2} = \sum_i \frac{1}{a} 
(\widetilde E_{x+\frac{a}{2} \hat i,i} - \widetilde E_{x-\frac{a}{2} \hat i,i}) + 
\frac{k}{2 \pi} B_x, \quad [\widetilde G_x,H] = 0, \quad 
x \in \widetilde \Lambda.
\end{eqnarray}
Note that $G_x$ has a different dimension than $G(x)$ in the continuum. In 
general, generators of gauge transformations at different sites of the same 
sublattice commute with each other, i.e.\ $[G_x,G_y] = 0$, 
$[\widetilde G_x,\widetilde G_y] = 0$. Let us also compute the commutation 
relations of the generator of a gauge transformation on the original lattice 
and a neighboring site on the dual lattice (cf.\ Fig.\ref{Gaussconsistency})
\begin{eqnarray}
[G_x,\widetilde G_{x+\frac{a}{2}\hat 1+\frac{a}{2}\hat 2}]&=&
- a^2 [E_{x+\frac{a}{2}\hat 1,1},\widetilde E_{x+\frac{a}{2}\hat 1,2}] 
- a^2 [E_{x+\frac{a}{2}\hat 2,2},\widetilde E_{x+\frac{a}{2}\hat 2,1}]
\nonumber \\ 
&+&\frac{k a^3}{2 \pi} [E_{x+\frac{a}{2}\hat 1,1},B_{x+\frac{a}{2}\hat 1+\frac{a}{2}\hat 2}]
+ \frac{k a^3}{2 \pi} [E_{x+\frac{a}{2}\hat 2,2},B_{x+\frac{a}{2}\hat 1+\frac{a}{2}\hat 2}] 
\nonumber \\
&+&\frac{k a^3}{2 \pi} [\widetilde E_{x+\frac{a}{2}\hat 2,1},\widetilde B_x] + 
\frac{k a^3}{2 \pi} [\widetilde E_{x+\frac{a}{2}\hat 1,2},\widetilde B_x] = 0.
\end{eqnarray}
Hence, generators of gauge transformations at different sites always commute.

The vacuum state $|0\rangle$ is gauge invariant and satisfies the Gauss law
$G_x |0\rangle = \widetilde G_x |0\rangle = 0$. Upon quantization, the classical
plane wave states turn into the states of two ``photons'', each of mass
$M = \frac{k e^2}{2 \pi}$, with the lattice dispersion relation 
$E(p) = \sqrt{M^2 + \hat p^2}$, $\hat p_i = \frac{2}{a} \sin\frac{p_i a}{2}$. A 
state $|Q,\widetilde Q\rangle$ that contains static charges $Q_x \in \Z$ (with 
$x \in \Lambda$) and static dual charges $\widetilde Q_x \in \Z$ (with 
$x \in \widetilde \Lambda$) satisfies 
\begin{equation}
\label{chargedstate}
G_x |Q,\widetilde Q\rangle = Q_x |Q,\widetilde Q\rangle, \, x \in \Lambda, 
\quad
\widetilde G_x |Q,\widetilde Q\rangle = \widetilde Q_x |Q,\widetilde Q\rangle, 
\, x \in \widetilde \Lambda.
\end{equation}

\subsection{Mutual Statistics of Original and Dual Charges}

Let us consider Wilson's parallel transporter $U_{x,i} = \exp(i a A_{x,i})$, which
moves a charge from $x-\frac{a}{2}\hat i$ to $x+\frac{a}{2}\hat i$. Under the 
unitary transformation
\begin{equation}
V = \prod_{x \in \Lambda} V_x = \prod_{x \in \Lambda} \exp(i \chi_x G_x),
\end{equation}
which implements a general gauge transformation in Hilbert space, the parallel 
transporter transforms as
\begin{equation}
U'_{x,i} = V U_{x,i} V^\dagger = 
\exp(i \chi_{x-\frac{a}{2}\hat i}) U_{x,i} \exp(- i \chi_{x+\frac{a}{2}\hat i}), \quad
x \in X, \widetilde X.
\end{equation}
The corresponding commutation relation takes the form
\begin{equation}
[G_y,U_{x,i}] = \frac{1}{a} \sum_j
[\Pi_{y+\frac{a}{2}\hat j,j} - \Pi_{y-\frac{a}{2}\hat j,j},U_{x,i}] = 
\left(\delta_{y+\frac{a}{2}\hat i,x} - \delta_{y-\frac{a}{2}\hat i,x}\right) U_{x,i}.
\end{equation}
Let us now act with $U_{x,i}$ on the charged state $|Q,\widetilde Q\rangle$, 
with external charges $Q_x$ at $x \in \Lambda$ and $\widetilde Q_x$ at 
$x \in \widetilde \Lambda$ (cf.\ eq.(\ref{chargedstate})), in order to 
investigate its effect on the charges
\begin{eqnarray}
G_y |Q',\widetilde Q\rangle&=&G_y U_{x,i} |Q,\widetilde Q\rangle = 
U_{x,i} G_y |Q,\widetilde Q\rangle + 
\left(\delta_{y+\frac{a}{2}\hat i,x} - \delta_{y-\frac{a}{2}\hat i,x}\right) U_{x,i} 
|Q,\widetilde Q\rangle \nonumber \\
&=&\left(Q_y + \delta_{y+\frac{a}{2}\hat i,x} - \delta_{y-\frac{a}{2}\hat i,x}\right)
|Q',\widetilde Q\rangle.
\end{eqnarray}
As expected $U_{x,i}$ acts as a charge transport operator which moves a charge 
 from $x-\frac{a}{2}\hat i$ to $x+\frac{a}{2}\hat i$, i.e.\ the state
$|Q',\widetilde Q\rangle = U_{x,i} |Q,\widetilde Q\rangle$ has charges
\begin{equation}
Q'_y = Q_y + \delta_{y+\frac{a}{2}\hat i,x} - \delta_{y-\frac{a}{2}\hat i,x}.
\end{equation}

\begin{figure}[tbp]
\begin{center}
\includegraphics[width=0.5\textwidth]{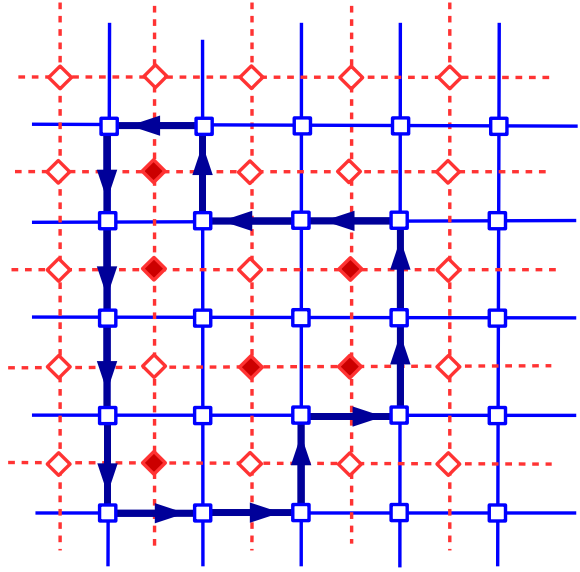}
\end{center}
\caption{[Color online] \textit{A Wilson loop $W_{\cal C}$ on the original 
lattice encircles a set of dual charges $\widetilde Q_x$ (filled diamonds). Up
to exponentially suppressed corrections due to the finite ``photon'' mass, the 
vacuum expectation value of the Wilson loop depends only on the total encircled 
dual charge, not on further details of the closed loop ${\cal C}$.}}
\label{Wilson}
\end{figure}
Let us now transport a charge around the closed loop ${\cal C}$ (on the
original lattice), so that it returns to its initial position (cf.\
Fig.\ref{Wilson}). The wave function then turns into
\begin{equation}
W_{\cal C} |Q,\widetilde Q\rangle = 
\prod_{(x,i) \in {\cal C}} U_{x,i} |Q,\widetilde Q\rangle =
\exp(i \Phi_{\cal C}) |Q,\widetilde Q\rangle,
\end{equation}
i.e., it picks up a phase 
\begin{equation}
\label{flux}
\Phi_{\cal C} = \sum_{x \in S \subset \widetilde \Lambda} a^2 B_x = 
\sum_{x \in S \subset \widetilde \Lambda} \frac{2 \pi}{k} \widetilde G_x -
\sum_{\cal C} \frac{2 \pi a}{k} \epsilon_{ij} \widetilde E_{x,j}, 
\end{equation}
which is given by the magnetic flux through the surface $S$ that is bounded
by ${\cal C} = \partial S$. Here $x$ denotes a point on the dual lattice at the
center of a plaquette on the original lattice which belongs to the surface $S$.
Eq.(\ref{flux}) results from Stoke's theorem on the lattice as well as from
eq.(\ref{gauge}). The first term on the right-hand side of eq.(\ref{flux}) 
counts the dual charges 
$\widetilde Q_S = \sum_{x \in S \subset \widetilde \Lambda} \widetilde G_x$
encircled by the loop ${\cal C}$. In a pure Chern-Simons theory (without a 
Maxwell term) this would be the only contribution. In that case, a charge 
that is transported around a closed loop ${\cal C}$ on the 
original lattice picks up a topological phase $\exp(2 \pi i \widetilde Q_S/k)$. 
In other words, charges on the original and on the dual lattice have mutual 
anyonic statistics, with the statistics angle $\frac{2 \pi}{k}$. The topological
phase is an adiabatic Berry phase. However, in contrast to generic
Berry phases, the topological phase $\Phi_{\cal C}$ is even insensitive to the 
shape of the curve ${\cal C}$, as long as it encircles the same total charge 
$\widetilde Q_S$ on the dual lattice. Obviously, corresponding rules about 
charge transport also apply to the dual lattice, with $U_{x,i}$ being replaced 
by $\widetilde U_{x,i} = \exp(i a \widetilde A_{x,i})$. In the 
Chern-Simons-Maxwell theory that we investigate here, a second term arises on
the right-hand side of eq.(\ref{flux}). This term is due to the ``photon'' cloud
surrounding the dual charges. Since the ``photon'' is massive, the effect of 
this term is exponentially suppressed at large distances \cite{Lue89}. Still, it
slightly affects the topological nature of the Berry phase and causes a certain
sensitivity to the precise location of the curve ${\cal C}$. The central result
of mutual anyonic statistics of original and dual charges with statistics angle
$\frac{2 \pi}{k}$ remains valid, as long as the charges are moved around each
other at distances much larger than the size of their massive ``photon'' clouds.

\section{Compact Chern-Simons-Maxwell Theory \\ on the Lattice}

In the previous section, we have considered a non-compact lattice theory with 
gauge group $\R$. As we have seen in eq.(\ref{crosscom}), the local degrees of 
freedom on a cross formed by a link and its dual link then have a ``mechanical''
analog, namely a charged ``particle'' moving in the 2-d plane $\R^2$ in the 
background of an abstract ``magnetic'' field $\frac{k}{2 \pi}$. Now we will 
compactify the gauge field and thus work with the gauge group $U(1)$. The 
``mechanical'' analog then corresponds to motion restricted to the torus 
$U(1)^2$. This leads to the quantization of the level $k$ in integer units, 
i.e.\ $k \in \Z$. As we will see, the electric field part of the Hamiltonian has
a 2-parameter family of self-adjoint extensions. Interestingly, fixing the two
self-adjoint extension parameters explicitly breaks the $U(1)$ gauge symmetry
down to $\Z(k)$.

\subsection{Commutation Relations of the Compact Theory}

In the compact theory, $A_{x,i}$ and $\widetilde A_{x,i}$ turn into phases of the 
parallel transporters
\begin{equation}
U_{x,i} = \exp(i a A_{x,i}) = \exp(i \varphi_{x,i}) \in U(1), \quad 
\widetilde U_{x,i} = \exp(i a \widetilde A_{x,i}) = 
\exp(i \widetilde \varphi_{x,i}) \in U(1),
\end{equation}
which now are the truly fundamental degrees of freedom, and
\begin{equation}
a E_{x,i}  = - i \partial_{\varphi_{x,i}} -
\frac{k}{4 \pi} \epsilon_{ij} \widetilde \varphi_{x,j}, \quad
a \widetilde E_{x,i}  = 
- i \partial_{\widetilde \varphi_{x,i}} -
\frac{k}{4 \pi} \epsilon_{ij} \varphi_{x,j}.
\end{equation}
Based on eq.(\ref{crosscom}), the local commutation relations now take the form
\begin{eqnarray}
&&[U_{x,i},U_{y,j}] = [\widetilde U_{x,i},\widetilde U_{y,j}] = 
[U_{x,i},\widetilde U_{y,j}] = 0, \nonumber \\
&&[E_{x,i},U_{y,j}] = \frac{1}{a} \delta_{ij} \delta_{xy} U_{x,i}, \quad
[\widetilde E_{x,i},\widetilde U_{y,j}] = 
\frac{1}{a} \delta_{ij} \delta_{xy} \widetilde U_{x,i}, \nonumber \\
&&[E_{x,i},\widetilde U_{y,j}] = [\widetilde E_{x,i},U_{y,j}] = 0, \nonumber \\
&&[E_{x,i},\widetilde E_{y,j}] = [\widetilde E_{x,i},E_{y,j}] = 
- i \frac{k}{2 \pi a^2} \epsilon_{ij} \delta_{xy}, \nonumber \\
&&[E_{x,i},E_{y,j}] = [\widetilde E_{x,i},\widetilde E_{y,j}] = 0.
\end{eqnarray}
Since, unlike $A_{x,i}$ and $\widetilde A_{x,i}$, $\varphi_{x,i}$ and 
$\widetilde \varphi_{x,i}$ are angles, they do not represent self-adjoint
quantum mechanical operators.

\subsection{Self-adjoint Extensions of the Electric Contribution to the 
Hamiltonian of a Single Cross}

Let us consider a single cross, formed by a link and its dual link. For 
concreteness, we consider a cross $x \in X$, i.e.\ the link variable 
$\varphi$ is associated with the 1-direction, while $\widetilde \varphi$ is
associated with the 2-direction. The situation for 
$x \in \widetilde X$ is completely analogous. In order to keep the notation
simple, in this subsection we suppress the link indices $(x,i)$, which are
uniquely determined for a specific cross $x \in X$. The ``mechanical'' analog of
the problem then corresponds to a charged ``particle'' with ``spatial'' 
coordinates $(\varphi,\widetilde \varphi)$ moving in the 2-dimensional group 
space subject to the cross-based Hamiltonian
\begin{equation}
H_+ = \frac{e^2 a^2}{2}(E_{x,1}^2 + \widetilde E_{x,2}^2) = 
\frac{e^2 a^2}{2}(E^2 + \widetilde E^2).
\end{equation}
It should be noted that, after compactification, a dual charge $\widetilde e$
can no longer be turned into $e$ by a field redefinition. In order to maintain 
the symmetry between the original and the dual lattice, we simply put 
$\widetilde e = e$. The electric field operators are now given by
\begin{equation}
a E = - i \partial_\varphi + a(\varphi,\widetilde \varphi), \quad
a \widetilde E = - i \partial_{\widetilde \varphi} +
\widetilde a(\varphi,\widetilde \varphi).
\end{equation}
Here $(a,\widetilde a)$ is an abstract vector potential on the group space 
torus $U(1)^2$ (pa\-ra\-me\-trized by $(\varphi,\widetilde \varphi)$) with
\begin{equation}
a(\varphi,\widetilde \varphi) = - \frac{k}{4 \pi} \widetilde \varphi, \quad
\widetilde a(\varphi,\widetilde \varphi) = \frac{k}{4 \pi} \varphi,
\end{equation}
which gives rise to the abstract ``magnetic'' field
\begin{equation}
b(\varphi,\widetilde \varphi) = 
\partial_\varphi \widetilde a(\varphi,\widetilde \varphi) -
\partial_{\widetilde \varphi} a(\varphi,\widetilde \varphi) = \frac{k}{2 \pi}.
\end{equation}
The level $k$ thus determines the value of the constant abstract ``magnetic''
field $b(\varphi,\widetilde \varphi)$ (which should not be confused with the
actual magnetic fields $B_x$ and $\widetilde B_x$).

Since the electric field part of the Hamiltonian $H_+$ contains $E^2$ and 
$\widetilde E^2$, these operators must be self-adjoint. For mathematical 
details related to the theory of self-adjoint extensions we refer to 
\cite{Ree72,Wei87}. The problem of finding the self-adjoint extensions of
$H_+$ has been discussed in 
\cite{AlH09} for the ``mechanical'' analog of a charged particle moving on a
torus in a constant magnetic field. The same mathematical solution applies 
here. The self-adjoint extension parameters enter the boundary conditions for 
the wave function $\Psi(\varphi,\widetilde \varphi)$ on the group space torus 
$U(1)^2$. This wave function will extend to the wave functional of the lattice 
field theory, once we combine all crosses to form the entire lattice. For the 
moment, we continue to consider a single cross in isolation.

While the abstract ``magnetic'' field $b$ is constant, and thus obviously 
periodic over the group space torus, the abstract vector potential 
$(a,\widetilde a)$ obeys the boundary conditions
\begin{eqnarray}
\label{torusbc}
&&a(\varphi + 2 \pi,\widetilde \varphi) = 
- \frac{k}{4 \pi} \widetilde \varphi = a(\varphi,\widetilde \varphi), 
\nonumber \\
&&a(\varphi,\widetilde \varphi + 2 \pi) = 
- \frac{k}{4 \pi} \widetilde \varphi - \frac{k}{2} =  
a(\varphi,\widetilde \varphi) - \frac{k}{2}, \nonumber \\
&&\widetilde a(\varphi + 2 \pi,\widetilde \varphi) =
\frac{k}{4 \pi} \varphi + \frac{k}{2} =
\widetilde a(\varphi,\widetilde \varphi) + \frac{k}{2}, \nonumber \\
&&\widetilde a(\varphi,\widetilde \varphi + 2 \pi) = 
\frac{k}{4 \pi} \varphi =  \widetilde a(\varphi,\widetilde \varphi).
\end{eqnarray}
Let us introduce the transition functions
\begin{equation}
\alpha(\widetilde \varphi) = - \frac{k}{2} \widetilde \varphi + 
\widetilde \theta, 
\quad
\widetilde \alpha(\varphi) = \frac{k}{2} \varphi + \theta,
\end{equation}
where $\theta, \widetilde \theta \in [0,2 \pi[$ will turn out to be two 
self-adjoint extension parameters. We can then express the boundary conditions 
of eq.(\ref{torusbc}) as twisted periodic boundary conditions over the group
space torus $U(1)^2$
\begin{eqnarray}
a(\varphi + 2 \pi,\widetilde \varphi) = a(\varphi,\widetilde \varphi) 
- \partial_\varphi \alpha(\widetilde \varphi), \nonumber \\
a(\varphi,\widetilde \varphi + 2 \pi) = a(\varphi,\widetilde \varphi) 
- \partial_\varphi \widetilde \alpha(\varphi), \nonumber \\
\widetilde a(\varphi + 2 \pi,\widetilde \varphi) = 
\widetilde a(\varphi,\widetilde \varphi) 
- \partial_{\widetilde \varphi} \alpha(\widetilde \varphi), \nonumber \\
\widetilde a(\varphi,\widetilde \varphi + 2 \pi) = 
\widetilde a(\varphi,\widetilde \varphi) 
- \partial_{\widetilde \varphi} \widetilde \alpha(\varphi).
\end{eqnarray}
As we see, $\alpha$ and $\widetilde \alpha$ play the role of 
abstract gauge transformations (not to be confused with the original lattice 
gauge symmetry). In order to respect the abstract gauge structure, the wave 
function $\Psi$ must obey consistent twisted periodic boundary conditions, i.e.
\begin{eqnarray}
\label{selfadj}
&&\Psi(\varphi + 2 \pi,\widetilde \varphi) =
\exp(i \alpha(\widetilde \varphi)) \Psi(\varphi,\widetilde \varphi) =
\exp\left(- i \frac{k}{2} \widetilde \varphi + i \widetilde \theta\right)
\Psi(\varphi,\widetilde \varphi), \nonumber \\
&&\Psi(\varphi,\widetilde \varphi + 2 \pi) =
\exp(i \widetilde \alpha(\varphi)) 
\Psi(\varphi,\widetilde \varphi) = 
\exp\left(i \frac{k}{2} \varphi + i \theta\right) 
\Psi(\varphi,\widetilde \varphi).
\end{eqnarray}
Note that this equation holds for a cross $x \in X$, while for 
$x \in \widetilde X$ the $\theta$ and $\widetilde \theta$ terms change sign.
Now we see explicitly, that $\theta$ and $\widetilde \theta$ enter the boundary
condition of the wave function, which is typical for self-adjoint extension
parameters. The above relations immediately imply
\begin{eqnarray}
\Psi(\varphi + 2 \pi,\widetilde \varphi + 2 \pi)&=&
\exp\left(- i \frac{k}{2}(\widetilde \varphi + 2 \pi) + 
i \widetilde \theta\right) \Psi(\varphi,\widetilde \varphi + 2 \pi) \nonumber \\
&=&\exp\left(- i \frac{k}{2}(\widetilde \varphi + 2 \pi) + i \widetilde \theta 
+ i \frac{k}{2} \varphi + i \theta\right) 
\Psi(\varphi,\widetilde \varphi), \nonumber \\
\Psi(\varphi + 2 \pi,\widetilde \varphi + 2 \pi)&=&
\exp\left(i \frac{k}{2}(\varphi + 2 \pi) + i \theta\right)  
\Psi(\varphi + 2 \pi,\widetilde \varphi) \nonumber \\
&=&\exp\left(i \frac{k}{2}(\varphi + 2 \pi) + i \theta 
- i \frac{k}{2} \widetilde \varphi + i \widetilde \theta\right) 
\Psi(\varphi,\widetilde \varphi).
\end{eqnarray}
Consistency of the two expressions requires $\exp(2 \pi i k) = 1$, which leads 
to the quantization of the level $k \in \Z$. This quantization condition implies
that the abstract ``magnetic'' flux through the group space torus $U(1)^2$ is 
quantized as well
\begin{equation}
\int_0^{2 \pi} d\varphi \int_0^{2 \pi} d\widetilde \varphi \
b(\varphi,\widetilde \varphi) = 2 \pi k.
\end{equation}
The quantization of the level $k$ thus manifests itself as a Dirac quantization
condition of the abstract ``magnetic'' flux that threads the group space torus.

Besides the field strength $b$, the abstract gauge field $(a,\widetilde a)$ on 
the group space torus $U(1)^2$ has two additional gauge invariant quantities 
--- the Polyakov loops that arise due to the non-trivial holonomies of the 
torus
\begin{eqnarray}
&&\phi(\widetilde \varphi) = 
\int_0^{2 \pi} d\varphi \ a(\varphi,\widetilde \varphi) +
\alpha(\widetilde \varphi) = - k \widetilde \varphi + \widetilde \theta, 
\nonumber \\
&&\widetilde \phi(\varphi) = 
\int_0^{2 \pi} d\widetilde \varphi \ \widetilde a(\varphi,\widetilde \varphi) + 
\widetilde \alpha(\varphi) = k \varphi + \theta.
\end{eqnarray}
Unlike the classical theory, the quantum theory is sensitive to the complex 
Aharonov-Bohm phases defined by the Polyakov loops
\begin{equation}
\exp(i \phi(\widetilde \varphi)) = 
\exp(- i k \widetilde \varphi + i \widetilde \theta), \quad
\exp(i \widetilde \phi(\varphi)) = \exp(i k \varphi + i \theta).
\end{equation}
Remarkably, the Polyakov loops explicitly break the continuous $U(1)^2$
translation invariance of the torus down to the discrete translation group
$\Z(k)^2$, since the corresponding Aharonov-Bohm phases are invariant only
under shifts of $\varphi$ and $\widetilde \varphi$ by integer multiples of
$\frac{2 \pi}{k}$. This corresponds to an explicit breaking of the original and 
dual lattice $U(1)$ gauge symmetries down to $\Z(k)$.

\subsection{Fate of the Original and Dual $U(1)$ Gauge Symmetries}

As we have just seen, when we fix the values of the self-adjoint extension
parameters $\theta$ and $\widetilde \theta$, the original and dual continuous 
$U(1)$ gauge symmetries of the classical compact doubled Chern-Simons-Maxwell 
theory are explicitly broken down to discrete $\Z(k)$ gauge symmetries. Since 
gauge symmetries just reflect redundancies in our
theoretical description and are not actual physical symmetries, one may wonder 
what happened at the quantum level to the redundancy associated with the 
continuous $U(1)$ gauge symmetries of the classical theory. In order to answer 
this question, let us perform original and dual continuous $U(1)$ gauge 
transformations on the link variables
\begin{eqnarray}
\label{Ugt}
U_{x,i}'&=&\exp(i \varphi_{x,i}') = 
\exp(i \chi_{x-\frac{a}{2}\hat i}) U_{x,i} \exp(- i \chi_{x+\frac{a}{2}\hat i})
\nonumber \\
&=&\exp(i (\varphi_{x,i} - \chi_{x+\frac{a}{2}\hat i} + \chi_{x-\frac{a}{2}\hat i})), 
\nonumber \\
\widetilde U_{x,i}'&=&\exp(i \widetilde \varphi_{x,i}') = 
\exp(i \widetilde \chi_{x-\frac{a}{2}\hat i}) \widetilde U_{x,i} 
\exp(- i \widetilde \chi_{x+\frac{a}{2}\hat i}) \nonumber \\
&=&\exp(i (\widetilde \varphi_{x,i} - 
\widetilde \chi_{x+\frac{a}{2}\hat i} + \widetilde \chi_{x-\frac{a}{2}\hat i})).
\end{eqnarray}
While the magnetic part of the Hamiltonian is by construction manifestly gauge
invariant, the electric part is at least not obviously gauge invariant, because 
$E$ and $\widetilde E$ depend on $\varphi$ and $\widetilde \varphi$. Still, in 
the non-compact theory, $E$ and $\widetilde E$ turn out to be gauge invariant, 
i.e.\ they commute with local gauge transformations $G$ and $\widetilde G$. Let 
us return to the compact theory and consider a general finite gauge 
transformation that is represented in Hilbert space by a unitary transformation
\begin{equation}
V = \prod_{x \in \Lambda} \exp(i \chi_x G_x) 
\prod_{x \in \widetilde \Lambda} \exp(i \widetilde \chi_x \widetilde G_x).
\end{equation}
It is straightforward to convince oneself that indeed
\begin{equation}
V U_{x,i} V^\dagger =  U_{x,i}', \quad 
V \widetilde U_{x,i} V^\dagger =  \widetilde U_{x,i}', 
\end{equation}
with $U_{x,i}'$ and $\widetilde U_{x,i}'$ given by eq.(\ref{Ugt}). Furthermore,
at least at a formal level,
\begin{equation}
V E_{x,i} V^\dagger =  E_{x,i}, \quad 
V \widetilde E_{x,i} V^\dagger =  \widetilde E_{x,i}, \quad
V H V^\dagger = H,
\end{equation}
which seems to suggest that the compact theory is still $U(1)$ gauge invariant.
However, this is not the case, because $V$ does not leave the domain of the
Hamiltonian invariant. In other words, the gauge transformed wave functional
$\Psi'[\varphi,\widetilde \varphi] = V \Psi[\varphi,\widetilde \varphi]$ no 
longer satisfies the self-adjoint extension condition eq.(\ref{selfadj}). 
Unlike the wave function $\Psi(\varphi,\widetilde \varphi)$ (which only depends 
on the link variables $\varphi$ and $\widetilde \varphi$ of a single cross),
the wave functional $\Psi[\varphi,\widetilde \varphi]$ depends on the entire 
lattice gauge fields $[\varphi,\widetilde \varphi]$. Under a general gauge 
transformation, one obtains
\begin{equation}
\Psi'[\varphi,\widetilde \varphi] = V \Psi[\varphi,\widetilde \varphi] =
\prod_{x \in \Lambda} \exp\left(i \frac{k a^2}{4 \pi} \chi_x \widetilde B_x\right)
\prod_{x \in \widetilde \Lambda} \exp\left(i \frac{k a^2}{4 \pi} 
\widetilde \chi_x B_x\right) \Psi[\varphi',\widetilde \varphi'],
\end{equation}
where $\varphi'$ and $\widetilde \varphi'$ are the gauge transformed fields of
eq.(\ref{Ugt}), and the plaquette magnetic fields are given by
\begin{eqnarray}
&&\exp(i a^2 B_x) = 
\exp(i (\varphi_{x-\frac{a}{2}\hat 2,1} + \varphi_{x+\frac{a}{2}\hat 1,2} 
- \varphi_{x+\frac{a}{2}\hat 2,1} - \varphi_{x-\frac{a}{2}\hat 1,2})), 
\quad x \in \widetilde \Lambda, \nonumber \\
&&\exp(i a^2 \widetilde B_x) = \exp(i (\widetilde \varphi_{x-\frac{a}{2}\hat 2,1} +
\widetilde \varphi_{x+\frac{a}{2}\hat 1,2} - \widetilde \varphi_{x+\frac{a}{2}\hat 2,1} -
\widetilde \varphi_{x-\frac{a}{2}\hat 1,2})), \quad x \in \Lambda.
\end{eqnarray}
When one applies the unitary transformation to the wave 
function of a single cross, one finds that the transformed wave function obeys 
the boundary condition
\begin{eqnarray}
&&\Psi'(\varphi + 2 \pi,\widetilde \varphi) =
\exp\left(- i \frac{k}{2} \widetilde \varphi + i \widetilde \theta'\right)
\Psi'(\varphi,\widetilde \varphi), \nonumber \\
&&\Psi'(\varphi,\widetilde \varphi + 2 \pi) =
\exp\left(i \frac{k}{2} \varphi + i \theta'\right) 
\Psi'(\varphi,\widetilde \varphi),
\end{eqnarray}
where the self-adjoint extension parameters $\theta$ and $\widetilde \theta$,
which can be associated with the link and its dual link, respectively, have 
been transformed to
\begin{eqnarray}
&&\exp(i \theta_{x,i}') = \exp(i (\theta_{x,i} + k \chi_{x+\frac{a}{2}\hat i} -
k \chi_{x-\frac{a}{2}\hat i})), \nonumber \\
&&\exp(i \widetilde \theta_{x,i}') = \exp(i (\widetilde \theta_{x,i} +
k \widetilde \chi_{x+\frac{a}{2}\hat i} - k \widetilde \chi_{x-\frac{a}{2}\hat i})).
\end{eqnarray}
This means that the self-adjoint extension parameters $\exp(i \theta_{x,i})$ and 
$\exp(i \widetilde \theta_{x,i})$ themselves also transform as $U(1)$ gauge 
fields. However, unlike $U_{x,i}$ and $\widetilde U_{x,i}$, they transport $k$ 
units of charge. As a consequence, $\exp(i \theta_{x,i})$ and 
$\exp(i \widetilde \theta_{x,i})$ are invariant against $\Z(k)$ gauge 
transformations. Since a $U(1)$ gauge transformation changes the self-adjoint
extension parameters $\theta$ and $\widetilde \theta$, and thus the domain of
the Hamiltonian, it can no longer be considered as a symmetry of the theory.
However, the redundancy associated with the $U(1)$ gauge symmetries of the 
classical theory is still present at the quantum level. Continuous $U(1)$ gauge 
transformations simply lead from one domain of $H$ (characterized by $\theta$
and $\widetilde \theta$) to a unitarily equivalent domain characterized by 
$\theta'$ and $\widetilde \theta'$. The gauge invariant physical content of the 
theory is thus not determined by the link parameters $\theta_{x,i}$ and 
$\widetilde \theta_{x,i}$ themselves, but by their plaquette field strengths
\begin{eqnarray}
\label{thetaplaquette}
&&\exp(i \eta_x) = \exp(i (\theta_{x-\frac{a}{2}\hat 2,1} + 
\theta_{x+\frac{a}{2}\hat 1,2} - \theta_{x+\frac{a}{2}\hat 2,1} - 
\theta_{x-\frac{a}{2}\hat 1,2})), \quad x \in \widetilde \Lambda, \nonumber \\
&&\exp(i \widetilde \eta_x) = \exp(i (\widetilde \theta_{x-\frac{a}{2}\hat 2,1} +
\widetilde \theta_{x+\frac{a}{2}\hat 1,2} - \widetilde \theta_{x+\frac{a}{2}\hat 2,1} -
\widetilde \theta_{x-\frac{a}{2}\hat 1,2})),\quad x \in \Lambda.
\end{eqnarray}
A natural choice is $\eta_x = \widetilde \eta_x = 0$ for all $x$. Another
choice is $\eta_x = \widetilde \eta_x = \pi$. Both of these choices leave 
the charge conjugation symmetry C intact, while other choices break C 
explicitly. The parity symmetry P is already explicitly broken by the 
Chern-Simons term. It should be pointed out that, unlike $U_{x,i}$ and 
$\widetilde U_{x,i}$, $\exp(i \theta_{x,i})$ and $\exp(i \widetilde \theta_{x,i})$ 
are not dynamical gauge degrees of freedom, but parameters that define a 
super-selection sector. In this sense, they are analogous to the 
$\theta$ vacuum angle in 4-dimensional non-Abelian gauge theories. In 
particular, different values of $\theta$ define different vacuum sectors. For 
$\theta \neq 0,\pi$, in non-Abelian gauge theories the vacuum angle explicitly 
breaks the CP symmetry. 

What is the significance of the reduction of the manifest gauge symmetry from
$U(1)$ to $\Z(k)$ at the quantum level? In particular, should one interpret 
this explicit quantum mechanical symmetry breaking as an anomaly? It is 
important to note that the quantum theory is sensitive to external parameters
(the gauge fields $\theta$ and $\widetilde \theta$ representing Aharonov-Bohm
phases) which do not affect the dynamics at the classical level. Remarkably, in
this case the external parameters are $\Z(k)$ gauge invariant but change under 
$U(1)$ gauge transformations. 

\subsection{Local Magnetic Translation Group on a Single Cross}

As we have stressed before, the actual manifest local symmetry of the 
Hamiltonian in a given domain is reduced to $\Z(k)$. To investigate this 
symmetry, first let us again consider the purely electric Hamiltonian 
$H_+ = \frac{e^2 a^2}{2} (E^2 + \widetilde E^2)$, which commutes with the unitary
operators
\begin{eqnarray}
&&T = \exp\left(\frac{2 \pi i}{k} \left(- i \partial_\varphi + \frac{k}{4 \pi}
\widetilde \varphi - \frac{\widetilde \theta}{2 \pi}\right)\right) = 
\exp\left(\frac{2 \pi}{k} \partial_\varphi + 
\frac{i}{2} \widetilde \varphi - \frac{i}{k} \widetilde \theta\right), 
\nonumber \\
&&\widetilde T = \exp\left(\frac{2 \pi i}{k} 
\left(- i \partial_{\widetilde \varphi} - \frac{k}{4 \pi}\varphi - 
\frac{\theta}{2 \pi}\right)\right) = 
\exp\left(\frac{2 \pi}{k} \partial_{\widetilde \varphi} -
\frac{i}{2} \varphi - \frac{i}{k} \theta\right),
\end{eqnarray}
that induce translations by $\frac{2 \pi}{k}$ of the wave function (up to an 
abstract gauge transformation), i.e.
\begin{eqnarray}
&&T \Psi(\varphi,\widetilde \varphi) = 
\exp\left(i \frac{\widetilde \varphi}{2} - \frac{i}{k} \widetilde \theta\right)
\Psi(\varphi + \frac{2 \pi}{k},\widetilde \varphi), \nonumber \\
&&\widetilde T \Psi(\varphi,\widetilde \varphi) = 
\exp\left(- i \frac{\varphi}{2} - \frac{i}{k} \theta\right)
\Psi(\varphi,\widetilde \varphi + \frac{2 \pi}{k}).
\end{eqnarray}
It is straightforward to rewrite the boundary condition eq.(\ref{selfadj}) as
\begin{equation}
\label{conditionT}
T^k \Psi(\varphi,\widetilde \varphi) = \Psi(\varphi,\widetilde \varphi), \quad
\widetilde T^k \Psi(\varphi,\widetilde \varphi) = 
\Psi(\varphi,\widetilde \varphi).
\end{equation}
This immediately implies that $T$ and $\widetilde T$ respect the domain 
structure of the Hamiltonian, i.e., after acting with these operators, the
wave function still obeys the boundary condition eq.(\ref{selfadj}) that defines
the domain of $H_+$. The operators $T$ and $\widetilde T$ obey
\begin{equation}
\label{Tcom}
\widetilde T T = \exp\left(\frac{2 \pi i}{k}\right) T \widetilde T.
\end{equation}
These two operators generate a discrete group ${\cal G}$, known as the magnetic
translation group \cite{Zak64,Wal09}, which consists of the elements
\begin{equation}
g(n,\widetilde n,m) = \exp\left(\frac{2 \pi i m}{k}\right) 
\widetilde T^{\widetilde n} T^n, \quad n, \widetilde n, m \in \{0,1,\dots,k - 1\}.
\end{equation}
The group multiplication rule is given by
\begin{equation}
g(n,\widetilde n,m) g(n',\widetilde n',m') = 
g(n + n',\widetilde n + \widetilde n',m + m' - n \widetilde n').
\end{equation}
Here all summations are understood modulo $k$. The unit element is given by
\begin{equation}
\1 = g(0,0,0),
\end{equation}
and the elements
\begin{equation}
z_m = g(0,0,m) = \exp\left(\frac{2 \pi i m}{k}\right),
\end{equation}
form a cyclic Abelian subgroup $\Z(k) \subset {\cal G}$. The inverse of the
group element $g(n,\widetilde n,m)$ is given by
\begin{equation}
g(n,\widetilde n,m)^{-1} = g(- n,- \widetilde n,- m - n \widetilde n).
\end{equation}
This follows because
\begin{equation}
g(n,\widetilde n,m) g(- n,- \widetilde n,- m - n \widetilde n) = 
g(0,0,- n \widetilde n + n \widetilde n) = g(0,0,0) = \1.
\end{equation}
Let us now consider the conjugacy class of a group element 
$g(n,\widetilde n,m)$ that consists of the elements
\begin{eqnarray}
&&\!\!\!\!\!\!\!\!\!\!g(n',\widetilde n',m') g(n,\widetilde n,m) 
g(n',\widetilde n',m')^{-1} = \nonumber \\
&&\!\!\!\!\!\!\!\!\!\!g(n' + n,\widetilde n' + \widetilde n,
m' + m - n' \widetilde n) g(- n',- \widetilde n',- m' - n' \widetilde n') = 
\nonumber \\
&&\!\!\!\!\!\!\!\!\!\!g(n,\widetilde n,m - n'(\widetilde n + \widetilde n') + 
(n' + n) \widetilde n') = 
g(n,\widetilde n,m + n \widetilde n' - n' \widetilde n).
\end{eqnarray}
The elements $g(0,0,m) = z_m \in \Z(k)$ are conjugate only to themselves and 
thus form $k$ single-element conjugacy classes. Multiplication by a phase $z_m$ 
amounts to a $\Z(k)$ gauge transformation. The conjugacy classes hence
correspond to gauge equivalence classes. The elements $g(0,0,m) = z_m$ commute 
with all other elements and thus form the center $\Z(k)$ of the group 
${\cal G}$. Since the individual elements of the center form separate conjugacy 
classes, the center forms a normal subgroup which can be factored out. This
corresponds to identifying gauge equivalence classes. It should be pointed out 
that ${\cal G}$ is not the direct product $\Z(k) \times \Z(k) \times \Z(k)$. 
Since the quotient space ${\cal G}/\Z(k)$ is not a subgroup of ${\cal G}$, 
${\cal G}$ is not even a semi-direct product of $\Z(k) \times \Z(k)$ and 
$\Z(k)$, but just a particular central extension of $\Z(k) \times \Z(k)$ by the 
center subgroup $\Z(k)$.

\subsection{Spectrum of the Hamiltonian on a Single Cross}

Before we consider the full Hamiltonian of the entire lattice theory, let
us again consider the purely electric Hamiltonian $H_+$ of a single cross. It
commutes with both $T$ and $\widetilde T$, which, however, don't commute with
each other. Let us construct simultaneous eigenstates of $H_+$ and $T$. Since 
$T^k = \1$, the eigenvalues of $T$ are given by $\exp(2 \pi i l/k)$ with 
$l \in \{0,1,\dots,k - 1\}$, while the eigenvalues of $H_+$ are given by 
$E_n = M (n + \frac{1}{2})$ (where $n \in \{0,1,2,\dots\}$ denotes the 
Landau level \cite{Lan30,Ste08} in the ``mechanical'' analog, and $M$ is the 
``photon'' mass). Hence, we can construct simultaneous eigenstates 
$|n l\rangle$ such that
\begin{equation}
H_+ |n l\rangle = M \left(n + \frac{1}{2}\right) |n l\rangle, \
T |n l\rangle = \exp\left(\frac{2 \pi i l}{k}\right) |n l\rangle.
\end{equation}
As a consequence of eq.(\ref{Tcom}) one obtains
\begin{equation}
T \widetilde T|n l\rangle = \exp\left(- \frac{2 \pi i}{k}\right) \widetilde T T
|n l\rangle = \exp\left(\frac{2 \pi i (l - 1)}{k}\right) 
\widetilde T|n l\rangle,
\end{equation}
which implies
\begin{equation}
\widetilde T |n l\rangle = |n (l - 1)\rangle.
\end{equation}
Since $[\widetilde T,H] = 0$, the $k$ states $|n l\rangle$ with 
$l \in 0,1,\dots,k - 1$ are degenerate and form an irreducible representation 
of the magnetic translation group. If we restrict ourselves to the lowest Landau
level, $n = 0$, by sending $e \rightarrow \infty$ such that 
$M \rightarrow \infty$, the local Hilbert space of a cross is $k$-dimensional.
As we will see later, this is the limit in which the doubled 
Chern-Simons-Maxwell theory reduces to the $\Z(k)$ variant of the toric code.

In order to prepare for the derivation of the toric code in a later subsection, 
it is useful to explicitly construct the eigenfunctions $|n l\rangle$, which 
are given by
\begin{eqnarray}
\label{psinl}
\Psi^{\theta,\widetilde \theta}_{nl}(\varphi,\widetilde \varphi)&=&
\langle \varphi \widetilde \varphi|n l\rangle = \frac{1}{\sqrt{2 \pi}}
\sum_{m \in \Z} \psi_n\left(\widetilde \varphi - 2 \pi 
\left[m + \frac{l}{k} + \frac{\widetilde \theta}{ 2 \pi k}\right]\right) 
\nonumber \\
&\times&\exp\left(i (k \varphi + \theta) \left[m + \frac{l}{k} + 
\frac{\widetilde \theta}{2 \pi k}\right] - 
i \frac{k \varphi \widetilde \varphi}{4 \pi}\right).
\end{eqnarray}
Here 
\begin{equation}
\psi_n(\widetilde \varphi) = \sqrt[4]{\frac{k}{2 \pi}}
\frac{1}{\sqrt{2^n n! \sqrt{\pi}}} 
\exp\left(- \frac{k}{2 \pi} \frac{\widetilde \varphi^2}{2}\right) 
H_n\left(\sqrt{\frac{k}{2 \pi}} \widetilde \varphi\right),
\end{equation}
(with $H_n$ denoting the Hermite polynomials) are the eigenfunctions of a 
1-di\-men\-sio\-nal harmonic oscillator with characteristic ``frequency'' given 
by the ``photon'' mass $M = \frac{k e^2}{2 \pi}$. The corresponding energy 
eigenvalue is thus given by $E_n = M (n + \frac{1}{2})$. It is
straightforward to convince oneself that these wave functions indeed satisfy
the correct boundary conditions eq.(\ref{selfadj}) and are eigenstates of $H_+$.
The boundary conditions on the torus parametrized by the self-adjoint 
extension parameters $\theta$ and $\widetilde \theta$ are given by
\begin{eqnarray}
\label{thetabc}
&&\Psi^{\theta + 2 \pi,\widetilde \theta}_{nl}(\varphi,\widetilde \varphi) =
\exp\left(i \frac{2 \pi l + \widetilde \theta}{k}\right)
\Psi^{\theta,\widetilde \theta}_{nl}(\varphi,\widetilde \varphi) =
\Omega_{ll'} \Psi^{\theta,\widetilde \theta}_{nl'}(\varphi,\widetilde \varphi), 
\nonumber \\
&&\Psi^{\theta,\widetilde \theta + 2 \pi}_{nl}(\varphi,\widetilde \varphi) =
\Psi^{\theta,\widetilde \theta}_{n,l+1}(\varphi,\widetilde \varphi) =
\widetilde \Omega_{ll'} 
\Psi^{\theta,\widetilde \theta}_{nl'}(\varphi,\widetilde \varphi).
\end{eqnarray}
The boundary conditions of eq.(\ref{thetabc}) take the form of 't Hooft's 
$U(k)$ twisted boundary conditions \cite{tHo79,tHo81}. The transition functions 
$\Omega, \widetilde \Omega \in U(k)$ play the role of $U(k)$ gauge 
transformations. Their matrix elements are given by
\begin{equation}
\quad \Omega_{ll'} = 
\exp\left(i \frac{2 \pi l + \widetilde \theta}{k}\right) \delta_{ll'}, \quad
\widetilde \Omega_{ll'} = \delta_{l+1,l'},
\end{equation}
where $\delta_{l+1,l'}$ is understood modulo $k$. The transition functions
satisfy the cocycle consistency condition
\begin{equation}
\label{cocycle}
\Omega \widetilde \Omega = 
\exp\left(\frac{2 \pi i}{k}\right) \widetilde \Omega \Omega,
\end{equation}
which is characterized by the $U(k)$ gauge invariant twist factor 
$\exp(2 \pi i/k) \in \Z(k)$. The twist factor is an element of the center of 
$U(k)$. The operators $\Omega$ and $\widetilde \Omega$ obey the same
commutation relation as $\widetilde T$ and $T$ (cf.\ eq.(\ref{Tcom})) and thus
also generate a representation of the magnetic translation group. As 
$\widetilde \theta$ is increased by $2 \pi$, the $\Z(k)$ electric flux quantum 
number $l$ increases by 1. An adiabatic change of the self-adjoint extension 
parameters thus allows us to continuously interpolate between the different 
discrete $\Z(k)$ electric flux sectors. 

\subsection{$U(k)$ Berry Gauge Fields}

It is natural to ask how the wave functions on a single cross respond to an 
adiabatic change of $\theta$ and $\widetilde \theta$. Since $k$ states 
$|nl\rangle$ (with $l \in \{0,1,\dots,k-1\}$) are degenerate, this gives rise 
to an abstract $U(k)$ non-Abelian Berry gauge field on the torus $U(1)^2$ 
parametrized by $\theta$ and $\widetilde \theta$,
\begin{eqnarray}
\label{Berry}
g_{ll'}(\theta,\widetilde \theta)&=& 
\langle nl|W^\dagger \partial_\theta W|nl'\rangle = 0, \quad
W = \exp\left(- \frac{i}{2 \pi}(\theta \widetilde \varphi + 
\widetilde \theta \varphi)\right), \nonumber \\
\widetilde g_{ll'}(\theta,\widetilde \theta)&=&
\langle nl|W^\dagger \partial_{\widetilde \theta} W|nl'\rangle = 
i \frac{\theta}{2 \pi k} \delta_{ll'}.
\end{eqnarray}
In order to account for the fact that different values of $\theta$ and
$\widetilde \theta$ correspond to different domains of the Hamiltonian, it is 
important to include the unitary transformation $W$ in the definition of the 
Berry connection. The $g_{ll'}$ and $\widetilde g_{ll'}$ are elements of two 
$k \times k$ anti-Hermitean matrix-valued fields $g$ and $\widetilde g$ that 
play the role of abstract $U(k)$ Berry vector potentials. As we see, $g$ turns 
out to be Abelian, and $\widetilde g$ even vanishes, at least in the gauge that 
we have picked. The corresponding Berry field strength, i.e.\ the abstract 
``magnetic'' field, is given by
\begin{equation}
h(\theta,\widetilde \theta) =
\partial_\theta \widetilde g(\theta,\widetilde \theta) - 
\partial_{\widetilde \theta} g(\theta,\widetilde \theta) + 
[g(\theta,\widetilde \theta),\widetilde g(\theta,\widetilde \theta)], \quad
h_{ll'}(\theta,\widetilde \theta) = \frac{i}{2 \pi k} \delta_{ll'}.
\end{equation}
It is again Abelian and constant over the 2-dimensional torus parametrized by
$\theta$ and $\widetilde \theta$. Since the field strength is proportional to 
the unit-matrix, it is even invariant under general non-Abelian $U(k)$ gauge 
transformations (resulting from a basis change in the subspace of degenerate 
states $|nl\rangle$). This means that it actually reduces to an Abelian $U(1)$ 
Berry field strength. Still, since the field strength is non-zero, it can give 
rise to non-trivial Berry phases. However, since it is Abelian, its use for 
quantum information processing will be limited. The total ``magnetic'' 
flux that threads the torus is given by $\frac{2 \pi}{k}$. This is consistent 
with the twist factor $\exp(2 \pi i/k) \in \Z(k)$ of eq.(\ref{cocycle}) that 
characterizes 't Hooft's twisted boundary condition \cite{tHo79,tHo81}. 

Besides the Berry field strength, there are gauge invariant Berry Polyakov 
loops that result from the holonomies of the torus parametrized by the periodic
self-adjoint extension parameters $(\theta,\widetilde \theta)$. By integrating
the Berry connection along a straight line wrapping around the torus, we obtain
\begin{equation}
\int_0^{2 \pi} d\theta \ g_{ll}(\theta,\widetilde \theta) = 0, \quad
\int_0^{2 \pi} d\widetilde \theta \ \widetilde g_{ll}(\theta,\widetilde \theta) = 
i \frac{\theta}{k}.
\end{equation}
The Polyakov loop matrices $\Phi$, $\widetilde \Phi \in U(k)$, which describe 
parallel transport around the torus, also receive contributions from the 
transition functions $\Omega$ and $\widetilde \Omega$ such that
\begin{equation}
\label{Berryphase}
\Phi_{ll'}(\widetilde \theta) = \Omega_{ll'} =
\exp\left(i \frac{2 \pi l + \widetilde \theta}{k}\right) \delta_{ll'}, \quad 
\widetilde \Phi_{ll'}(\theta) = \exp\left(i \frac{\theta}{k} \right) 
\widetilde \Omega_{ll'} = \exp\left(i \frac{\theta}{k} \right) \delta_{l+1,l'}.
\end{equation}
As we will see later, this will allow us to derive the mutual statistics angle
$\frac{2 \pi}{k}$ also in the compact Chern-Simons theory.

\subsection{$\Z(k)$ Gauge Symmetry and Gauss Law}

The operators $T$ and $\widetilde T$ are associated with the links of a given
cross. Until now we have considered a cross centered at $x \in X$. Now
we also add the corresponding objects for $x \in \widetilde X$ such that
\begin{eqnarray}
&&T_{x,i} = \exp\left(\frac{2 \pi}{k} \partial_{\varphi_{x,i}} + 
\frac{i}{2} \epsilon_{ij} \widetilde \varphi_{x,j} -
\frac{i}{k} \epsilon_{ij} \widetilde \theta_{x,j}\right), \nonumber \\
&&\widetilde T_{x,i} = 
\exp\left(\frac{2 \pi}{k} \partial_{\widetilde \varphi_{x,i}} +
\frac{i}{2} \epsilon_{ij} \varphi_{x,j} +
\frac{i}{k} \epsilon_{ij} \theta_{x,j}\right).
\end{eqnarray}
A $\Z(k)$ Gauss law combines $T$ or $\widetilde T$ operators from links
that touch a common lattice point to form unitary operators that represent 
local $\Z(k)$ gauge transformations
\begin{equation}
V_x = \prod_i T_{x+\frac{a}{2}\hat i,i} T_{x-\frac{a}{2}\hat i,i}^\dagger, \, 
x \in \Lambda, \quad \widetilde V_x = \prod_i \widetilde T_{x+\frac{a}{2}\hat i,i} 
\widetilde T_{x-\frac{a}{2}\hat i,i}^\dagger, \, x \in \widetilde \Lambda.
\end{equation}
By construction, $V_x$ and $\widetilde V_x$ represent the manifest $\Z(k)$ gauge
symmetries of the full Hamiltonian (including both the electric cross and the
magnetic plaquette contributions), which respect the domain structure defined
by the values of the self-adjoint extension parameters $\theta$ and 
$\widetilde \theta$. Although $T$ and $\widetilde T$ on the same cross do not 
commute, original and dual gauge transformations on neighboring sites commute
(cf.\ Fig.\ref{Gaussconsistency})
\begin{eqnarray}
&&V_x \widetilde V_{x+\frac{a}{2}\hat 1+\frac{a}{2}\hat 2} = \nonumber \\
&&T_{x+\frac{a}{2}\hat 1,1} T_{x+\frac{a}{2}\hat 2,2} T_{x-\frac{a}{2}\hat 1,1}^\dagger
T_{x-\frac{a}{2}\hat 2,2}^\dagger \widetilde T_{x+a\hat 1+\frac{a}{2}\hat 2,1} 
\widetilde T_{x+\frac{a}{2}\hat 1+a\hat 2,2} \widetilde T_{x+\frac{a}{2}\hat 2,1}^\dagger 
\widetilde T_{x+\frac{a}{2}\hat 1,2}^\dagger = \nonumber \\
&&\widetilde T_{x+a\hat 1+\frac{a}{2}\hat 2,1} 
\widetilde T_{x+\frac{a}{2}\hat 1+a\hat 2,2} \widetilde T_{x+\frac{a}{2}\hat 2,1}^\dagger 
\widetilde T_{x+\frac{a}{2}\hat 1,2}^\dagger
T_{x+\frac{a}{2}\hat 1,1} T_{x+\frac{a}{2}\hat 2,2} T_{x-\frac{a}{2}\hat 1,1}^\dagger
T_{x-\frac{a}{2}\hat 2,2}^\dagger = \widetilde V_{x+\frac{a}{2}\hat 1+\frac{a}{2}\hat 2} V_x.
\nonumber \\ \,
\end{eqnarray}
Here we have used
\begin{eqnarray}
&&T_{x+\frac{a}{2}\hat 1,1} \widetilde T_{x+\frac{a}{2}\hat 1,2}^\dagger =
\exp\left(\frac{2 \pi i}{k}\right) \widetilde T_{x+\frac{a}{2}\hat 1,2}^\dagger
T_{x+\frac{a}{2}\hat 1,1}, \nonumber \\
&&T_{x+\frac{a}{2}\hat 2,2} \widetilde T_{x+\frac{a}{2}\hat 2,1}^\dagger = 
\exp\left(- \frac{2 \pi i}{k}\right) \widetilde T_{x+\frac{a}{2}\hat 2,1}^\dagger
T_{x+\frac{a}{2}\hat 2,2}.
\end{eqnarray}
For given $\theta$ and $\widetilde \theta$, physical states without external
charges from the corresponding domain of $H$ must obey the $\Z(k)$ Gauss law
\begin{equation}
V_x |\Psi\rangle = |\Psi\rangle, \, x \in \Lambda, \quad 
\widetilde V_x |\Psi\rangle = |\Psi\rangle, \, x \in \widetilde \Lambda.
\end{equation}
States that have external charges $Q_x, \widetilde Q_x \in {0,1,\dots,k-1}$ on 
the original and dual lattice obey
\begin{eqnarray}
\label{Zkcharges}
&&V_x |Q,\widetilde Q\rangle = \exp\left(\frac{2 \pi i}{k} Q_x\right)
|Q,\widetilde Q\rangle, \quad x \in \Lambda, \nonumber \\
&&\widetilde V_x |Q,\widetilde Q\rangle = 
\exp\left(\frac{2 \pi i}{k} \widetilde Q_x\right)|Q,\widetilde Q\rangle, \quad
x \in \widetilde \Lambda.
\end{eqnarray}
As we have seen before, $\theta$ and $\widetilde \theta$ represent external 
non-dynamical $U(1)$ lattice gauge fields. When we perform $U(1)$ gauge
transformations, we may change the values of $\theta$ and $\widetilde \theta$, 
which leads us into a unitarily equivalent domain of $H$, thus leaving the 
physics invariant. The physics does, however, depend on the gauge invariant
plaquette field strengths $\exp(i \eta_x)$ and $\exp(i \widetilde \eta_x)$
of eq.(\ref{thetaplaquette}). 

\subsection{Pure Chern-Simons Limit}

Let us now take the limit of large $e^2$, which implies that the ``photon'' of
mass $M = \frac{k e^2}{2 \pi}$ is removed from the spectrum and the system 
reduces to a pure Chern-Simons theory. In particular, the low-energy physics is 
restricted to the lowest Landau level $n=0$. Then there are only $k$ states 
left on each individual cross. As before, we choose an electric flux basis of 
the local Hilbert space, characterized by the fluxes $l \in \{0,1,\dots,k-1\}$ 
on the links of the original lattice. The electric field energy is then given
by $\frac{M}{2}$ for each cross, and the dynamics is entirely controlled by 
the magnetic field energy as well as by the Gauss law. Since they are suppressed
by $\frac{1}{e^2}$, the magnetic contributions to the energy can then be treated
in first order degenerate perturbation theory, by evaluating their matrix
elements between the degenerate electric flux eigenstates $|0l\rangle$. Putting 
$n=0$ in eq.(\ref{psinl}), it is straightforward to obtain
\begin{eqnarray}
\label{Tmatrices}
\langle 0 l|\exp(\pm i \varphi)|0 l'\rangle&=& 
C \exp\left(\pm \frac{i \theta}{k}\right) \delta_{l,l'\pm1} =
C \exp\left(\pm \frac{i \theta}{k}\right) 
\langle 0 l|\widetilde T^{\mp 1}|0 l'\rangle, \nonumber \\
\langle 0 l|\exp(\pm i \widetilde \varphi)|0 l'\rangle&=& 
C \exp\left(\pm \frac{i \widetilde \theta}{k}\right) 
\exp\left(\pm \frac{2 \pi i l}{k}\right) \delta_{ll'} \nonumber \\
&=&C \exp\left(\pm \frac{i \widetilde \theta}{k}\right) 
\langle 0 l|T^{\pm 1}|0 l'\rangle, \quad C = \exp\left(- \frac{\pi}{2 k}\right).
\end{eqnarray}
Using $\widetilde T|0l\rangle = |0(l-1)\rangle$, this implies that (up to a
trivial additive constant) the effective low-energy Hamiltonian (in the $n=0$ 
sector) is given by
\begin{eqnarray}
\label{toriccode}
H_{\rm eff}&=&- \frac{C^4}{2 e^2 a^2} \sum_{x \in \widetilde \Lambda} 
\left[\exp\left(\frac{i \eta_x}{k}\right) \widetilde V_x +
\exp\left(- \frac{i \eta_x}{k}\right) \widetilde V_x^\dagger\right] \nonumber \\
&-&\frac{C^4}{2 e^2 a^2} \sum_{x \in \Lambda} 
\left[\exp\left(\frac{i \widetilde \eta_x}{k}\right) V_x^\dagger + 
\exp\left(- \frac{i \widetilde \eta_x}{k}\right) V_x\right],
\end{eqnarray}
where $\eta$ and $\widetilde \eta$ are the plaquette field strength variables 
associated with the non-dynamical gauge fields $\theta$ and $\widetilde \theta$
of eq.(\ref{thetaplaquette}). The first contribution results from a 
$\cos(a^2 B_x)$ term on the original plaquettes and the second contribution 
originates from a dual plaquette term $\cos(a^2 \widetilde B_x)$. When the 
Hamiltonian acts on eigenstates $|Q,\widetilde Q\rangle$ with charges 
$Q_x, \widetilde Q_x$ on the original and on the dual lattice 
(cf.\ eq.(\ref{Zkcharges})) one obtains
\begin{equation}
H_{\rm eff}|Q,\widetilde Q\rangle = 
- \frac{C^4}{e^2 a^2} \left[\sum_{x \in \widetilde \Lambda} 
\cos\left(\frac{2 \pi \widetilde Q_x + \eta_x}{k}\right) + \sum_{x \in \Lambda}
\cos\left(\frac{2 \pi Q_x - \widetilde \eta_x}{k}\right)\right]
|Q,\widetilde Q\rangle.
\end{equation}
Hence, the effective Hamiltonian punishes violations of the $\Z(k)$ Gauss law 
on the original and on the dual lattice. In particular, for 
$\eta_x = \widetilde \eta_x = 0$, the ground state is reached when 
$V_x |\Psi\rangle = |\Psi\rangle$ for $x \in \Lambda$ and 
$\widetilde V_x |\Psi\rangle = |\Psi\rangle$ for $x \in \widetilde \Lambda$,
i.e.\ when $Q_x = \widetilde Q_x = 0$. For other values of $\eta_x$ and 
$\widetilde \eta_x$ other constellations of charges are energetically favored.

Let us construct the states $|Q,\widetilde Q\rangle$ more explicitly. For this
purpose, we construct dual charge projection operators at each site
$x \in \widetilde \Lambda$
\begin{equation}
P_{\widetilde Q_x} = \frac{1}{k} \sum_{m=0}^{k-1} 
\exp\left(- \frac{2 \pi i m}{k} \widetilde Q_x\right) \widetilde V_x^m,
\end{equation}
which obey
\begin{equation}
P_{\widetilde Q_x} P_{\widetilde Q_x'} = \delta_{\widetilde Q_x,\widetilde Q_x'}
P_{\widetilde Q_x}, \quad \widetilde V_x P_{\widetilde Q_x} = 
\exp\left(\frac{2 \pi i}{k} \widetilde Q_x\right) P_{\widetilde Q_x}.
\end{equation}
We now consider a state $|[l]\rangle$ corresponding to a specific configuration
$[l]$ of electric fluxes that obeys the Gauss law in the presence of charges 
$Q_x \in \Z(k)$ on the original lattice, $x \in \Lambda$, i.e.
\begin{equation}
\label{Gaussflux}
\sum_i (l_{x+\frac{a}{2}\hat i,i} - l_{x-\frac{a}{2}\hat i,i}) \mbox{mod} k = Q_x.
\end{equation}
By construction, we then have
\begin{equation}
V_x |[l]\rangle = \exp\left(\frac{2 \pi i}{k} Q_x\right) |[l]\rangle, \quad
x \in \Lambda.
\end{equation}
Next we apply the charge projection operators at all dual sites to obtain
\begin{equation}
|Q,\widetilde Q\rangle = {\cal N}
\prod_{x \in \widetilde \Lambda} P_{\widetilde Q_x} |[l]\rangle,
\end{equation}
where ${\cal N}$ is a normalization factor. By construction, this state indeed 
obeys
\begin{equation}
\widetilde V_x |Q,\widetilde Q\rangle = 
\exp\left(\frac{2 \pi i}{k} \widetilde Q_x\right)|Q,\widetilde Q\rangle, \quad x \in \widetilde\Lambda.
\end{equation}

As we have seen in Fig.\ref{Wilson}, in the non-compact theory the transport of
a charge along a curve ${\cal C} = \partial S$ on the original lattice, that 
encircles a total charge 
$\widetilde Q_S = \sum_{x \in S \subset \widetilde \Lambda} \widetilde Q_x$ on the dual
lattice, is associated with a non-trivial topological phase 
$\exp(2 \pi i \widetilde Q_S/k)$, thus showing that original and dual charges 
have mutually anyonic statistics with the statistics angle $\frac{2 \pi}{k}$. 
We will now derive the same result in the compact theory. For this purpose we 
again consider a Wilson loop
\begin{equation}
W_{\cal C} = \prod_{(x,i) \in {\cal C}} U_{x,i} = 
\prod_{(x,i) \in {\cal C}} \exp(i \varphi_{x,i}).
\end{equation}
Using eq.(\ref{Tmatrices}) we then obtain
\begin{equation}
W_{\cal C}|Q,\widetilde Q\rangle = C^{L_{\cal C}} 
\sum_{x \in S \subset \widetilde \Lambda} \widetilde V_x
\prod_{(x,i) \in {\cal C}} \exp(i \frac{\theta_{x,i}}{k}) |Q,\widetilde Q\rangle = 
C^{L_{\cal C}} 
\exp\left(\frac{2 \pi i \widetilde Q_S}{k} + i \frac{\eta_S}{k}\right)
|Q,\widetilde Q\rangle.
\end{equation}
Here $L_{\cal C}$ is the length of the closed loop ${\cal C}$ and
$\widetilde Q_S = \sum_{x \in S \subset \widetilde \Lambda} \widetilde Q_x$ is the
total dual charge encircled by ${\cal C}$. As before, the phase of the Wilson 
loop contains the $\widetilde Q_S$-term which implies that original and dual 
charges have mutually anyonic statistics with the statistics angle 
$\frac{2 \pi}{k}$. Due to the non-trivial background gauge field $\theta$, there
is an additional phase determined by the total background flux 
$\eta_S = \sum_{x \in S \subset \widetilde \Lambda} \eta_x$ encircled by ${\cal C}$.

The Wilson loop describes the instantaneous transport of a charge around the
loop ${\cal C}$. Alternatively, we now want to consider a much slower adiabatic 
process in which a charge-anti-charge pair is created, transported around the
loop ${\cal C}$, and finally annihilated. For this purpose, we compute the
Berry phase that we accumulate when we gradually change the background field
$\widetilde \theta$ from 0 to $2 \pi$ on all dual links intersecting the loop
${\cal C}$. According to eq.(\ref{Berryphase}) this provides us with a factor 
$\exp(i \theta_{x,i}/k)$ for all links that belong to ${\cal C}$. By Stokes 
theorem, this again yields the phase $\exp(i \eta_S/k)$. In addition, the factor
$\delta_{l+1,l'}$ in eq.(\ref{Berryphase}) shifts all electric flux variables 
along ${\cal C}$ by 1, which again gives rise to the phase 
$\exp(2 \pi i \widetilde Q_S/k)$ when one returns to the initial state 
$|Q,\widetilde Q\rangle$ at the end of the adiabatic charge transport process.
This shows explicitly that the anyon statistics angle $\frac{2 \pi}{k}$ is 
insensitive to the details of how the charge transport process is realized, be 
it adiabatic or instantaneous.

\subsection{Relation to the Toric Code}

The toric code is a $\Z(2)$ lattice gauge theory with a degenerate ground state 
that can be used as a storage device for quantum information that is 
topologically protected against decoherence. The fundamental degrees of freedom 
on each link of a square lattice are quantum spins $\frac{1}{2}$. In fact, the
toric code is a simple member of a large class of unconventional lattice gauge
theories known as quantum link models \cite{Hor81,Orl90,Cha97,Bro99,Bro04}. The 
Hamiltonian of the $\Z(2)$ toric code is given by
\begin{equation}
H = - J \sum_{x \in \widetilde \Lambda} U_x - G \sum_{x \in \Lambda} V_x, \quad
V_x = \prod_i \exp\left(i \pi S^3_{x+\frac{a}{2}\hat i}\right) 
\exp\left(- i \pi S^3_{x-\frac{a}{2}\hat i}\right).
\end{equation}
There is a spin $\frac{1}{2}$, $\vec S_{x,i}$, associated with each link 
connecting the neighboring lattice sites $x-\frac{a}{2}\hat i$ and
$x+\frac{a}{2}\hat i$. The quantum link operator is given by 
$U_{x,i} = S_{x,i}^1$, such that the plaquette term 
\begin{equation}
U_x = S_{x-\frac{a}{2}\hat 2,1}^1 S_{x+\frac{a}{2}\hat 1,2}^1
S_{x+\frac{a}{2}\hat 2,1}^1 S_{x-\frac{a}{2}\hat 1,2}^1, \quad x \in \widetilde \Lambda,
\end{equation}
is invariant under discrete $\Z(2)$ gauge transformations $\Omega_x = \pm 1$. 
The $G$-term with $G > 0$ punishes violations of the $\Z(2)$ Gauss law 
$V_x |\Psi\rangle = |\Psi\rangle$. Note that for $\Z(2)$ one has 
$U_x = U_x^\dagger$ and $V_x = V_x^\dagger$.

One can describe the dynamics of the toric code in a basis of electric flux
states on each link, which are characterized by the 3-component 
$S_{x,i}^3 = \pm \frac{1}{2}$ of the spin. The two spin values then correspond 
to two directions of the electric flux on each link. The quantum link operator 
$U_{x,i} = S_{x,i}^1$ reverses the direction of the electric flux, and thus the
plaquette operator $U_x$ reverses the direction of all electric fluxes
encircling the plaquette. The $\Z(2)$ toric code naturally generalizes to a
$\Z(k)$ variant with $k$ distinct values of the electric flux on each link. The
corresponding quantum link operator then increases the flux by one unit modulo
$k$. The $\Z(k)$ Gauss law ensures that the fluxes entering a common lattice
point add up to zero (again modulo $k$). Interestingly, the $\Z(k)$ variant of 
the toric code is exactly what emerged from the doubled compact 
Chern-Simons-Maxwell theory in the pure Chern-Simons limit of large ``photon'' 
mass, when we identify $J = G = \frac{C^4}{e^2}$.

\subsection{Rotating $\theta$ and $\widetilde \theta$ into the Hamiltonian}

Until now, the external background gauge fields $\theta$ and 
$\widetilde \theta$ entered the theory via the boundary conditions 
eq.(\ref{selfadj}) that define the domain of the Hamiltonian. For completeness 
we would now like to show that the background fields can be rotated into the 
Hamiltonian by the unitary transformation
\begin{eqnarray}
&&W = \prod_{x \in X} 
\exp\left(- \frac{i}{2 \pi}(\theta_{x,1} \widetilde \varphi_{x,2}
+ \widetilde \theta_{x,2} \varphi_{x,1}) \right)
\prod_{x \in \widetilde X}
\exp\left(\frac{i}{2 \pi}(\theta_{x,2} \widetilde \varphi_{x,1}
+ \widetilde \theta_{x,1} \varphi_{x,2}) \right), \nonumber \\
&&H' = W H W^\dagger, \quad |\Psi'\rangle = W |\Psi\rangle.
\end{eqnarray}
The original Hamiltonian
\begin{eqnarray}
H&=&\frac{e^2}{2} \sum_{x \in X} \left[\left(- i \partial_{\varphi_{x,1}} - 
\frac{k}{4 \pi} \widetilde \varphi_{x,2}\right)^2 +
\left(- i \partial_{\widetilde \varphi_{x,2}} + \frac{k}{4 \pi} \varphi_{x,1}
\right)^2\right] \nonumber \\
&+&\frac{e^2}{2} \sum_{x \in \widetilde X} \left[\left(- i \partial_{\varphi_{x,2}} + 
\frac{k}{4 \pi} \widetilde \varphi_{x,1}\right)^2 +
\left(- i \partial_{\widetilde \varphi_{x,1}} - \frac{k}{4 \pi} \varphi_{x,2}
\right)^2\right] \nonumber \\
&-&\frac{1}{e^2 a^2} \sum_{x \in \Lambda} \cos(a^2 \widetilde B_x) -
\frac{1}{e^2 a^2} \sum_{x \in \widetilde \Lambda} \cos(a^2 B_x),
\end{eqnarray}
with the boundary condition (for a cross with $x \in X$)
\begin{equation}
\Psi(\varphi + 2 \pi,\widetilde \varphi) =
\exp\left(- i \frac{k}{2} \widetilde \varphi + i \widetilde \theta\right)
\Psi(\varphi,\widetilde \varphi), \,
\Psi(\varphi,\widetilde \varphi + 2 \pi) =
\exp\left(i \frac{k}{2} \varphi + i \theta\right) 
\Psi(\varphi,\widetilde \varphi),
\end{equation}
then turns into the new Hamiltonian
\begin{eqnarray}
H'&=&\frac{e^2}{2} \sum_{x \in X} \left[\left(- i \partial_{\varphi_{x,1}} - 
\frac{k}{4 \pi} \widetilde \varphi_{x,2} +
\frac{\widetilde \theta_{x,2}}{2 \pi} \right)^2 +
\left(- i \partial_{\widetilde \varphi_{x,2}} + \frac{k}{4 \pi} \varphi_{x,1} +
\frac{\theta_{x,1}}{2 \pi} \right)^2\right] \nonumber \\
&+&\frac{e^2}{2} \sum_{x \in \widetilde X} \left[\left(- i \partial_{\varphi_{x,2}} + 
\frac{k}{4 \pi} \widetilde \varphi_{x,1} -
\frac{\widetilde \theta_{x,1}}{2 \pi} \right)^2 +
\left(- i \partial_{\widetilde \varphi_{x,1}} - \frac{k}{4 \pi} \varphi_{x,2} -
\frac{\theta_{x,2}}{2 \pi} \right)^2\right] \nonumber \\
&-&\frac{1}{e^2 a^2} \sum_{x \in \Lambda} \cos(a^2 \widetilde B_x) -
\frac{1}{e^2 a^2} \sum_{x \in \widetilde \Lambda} \cos(a^2 B_x),
\end{eqnarray}
The new Hamiltonian depends on $\theta$ and $\widetilde \theta$, while the new 
wave function now obeys the simplified boundary condition (again for a cross 
with $x \in X$)
\begin{equation}
\Psi'(\varphi + 2 \pi,\widetilde \varphi) =
\exp\left(- i \frac{k}{2} \widetilde \varphi\right)
\Psi'(\varphi,\widetilde \varphi), \quad
\Psi'(\varphi,\widetilde \varphi + 2 \pi) =
\exp\left(i \frac{k}{2} \varphi\right) \Psi'(\varphi,\widetilde \varphi).
\end{equation}
which is $\theta$- and $\widetilde \theta$-independent. The wave function on
a single cross $x \in X$ then takes the form
\begin{eqnarray}
{\Psi'}^{\theta,\widetilde \theta}_{nl}(\varphi,\widetilde \varphi)&=&
\langle \varphi \widetilde \varphi|n l\rangle =  \frac{1}{\sqrt{2 \pi}}
\sum_{m \in \Z} \psi_n\left(\widetilde \varphi - 2 \pi 
\left[m + \frac{l}{k} + \frac{\widetilde \theta}{ 2 \pi k}\right]\right) 
\nonumber \\
&\times&\exp\left(i \varphi\left[k m + l -
\frac{k \widetilde \varphi}{4 \pi}\right] + 
i \left[m + \frac{l}{k} + \frac{\widetilde \theta}{2 \pi k} -
\frac{\widetilde \varphi}{2 \pi}\right]\theta\right).
\nonumber \\ \,
\end{eqnarray}
In this case, in contrast to eq.(\ref{Berry}), and using 
$|nl\rangle' = W |nl\rangle$, the definition of the Berry connection no longer 
requires the insertion of $W$ and is simply given by
\begin{equation}
g_{ll'}(\theta,\widetilde \theta) =
{'\langle nl|\partial_\theta |nl'\rangle'}, \quad 
\widetilde g_{ll'}(\theta,\widetilde \theta) = 
{'\langle nl|\partial_{\widetilde \theta} |nl'\rangle'}.
\end{equation}
The resulting values of the Berry connection are the same as in 
eq.(\ref{Berry}).

\section{Conclusions}

We have obtained the $\Z(k)$ variant of the toric code in the infinite 
``photon'' mass limit of a doubled $U(1)$ Chern-Simons-Maxwell lattice gauge
theory. In contrast to ordinary lattice gauge theories, in this case the field
algebra is not link-based. Instead it is based on a cross formed by a link and
its corresponding dual link. In ordinary lattice gauge theory, each individual
link has a ``mechanical'' analog: a ``particle'' moving in the group space, 
i.e.\ a ``particle'' moving on a circle for a compact Abelian $U(1)$ lattice 
gauge theory. In the cross-based doubled compact Chern-Simons-Maxwell theory, 
on the other hand, the ``mechanical'' analog is a charged ``particle'' moving 
on a torus $U(1)^2$ which is threaded by an abstract ``magnetic'' flux, 
determined by the prefactor $k$ of the Chern-Simons term. The Dirac 
quantization condition for the abstract ``magnetic'' flux then leads to the 
quantization of the level $k \in \Z$.

Remarkably, the cross-based electric contribution to the Hamiltonian has two
self-adjoint extension parameters $\theta$ and $\widetilde \theta$, which 
themselves form two non-dynamical background $U(1)$ lattice gauge fields. In a
fixed background, the manifest gauge symmetry of the compact 
Chern-Simons-Maxwell theory is then reduced from $U(1)$ to $\Z(k)$. In 
particular, only a $\Z(k)$ (and not the full $U(1)$) Gauss law must be imposed
on physical states in a given domain of the Hamiltonian (which is determined
by $\theta$ and $\widetilde \theta$). The explicit symmetry breaking from 
$U(1)$ to $\Z(k)$ has a quantum mechanical origin, because only the quantum but 
not the classical theory is sensitive to the external self-adjoint extension 
parameters, which manifest themselves as Aharonov-Bohm phases. One might thus 
describe the quantum mechanical gauge symmetry breaking as an ``anomaly''. 
However, it is more like the explicit breaking of CP invariance due to a 
non-zero $\theta$-vacuum angle in 4-dimensional non-Abelian gauge theories. It 
is remarkable that a gauge symmetry can be broken by quantum effects in a 
similar manner. It should, however, be pointed out that the redundancy 
associated with the $U(1)$ gauge symmetry of the classical theory still 
persists at the quantum level. It manifests itself in the $U(1)$ gauge 
redundancy of the external background lattice gauge fields formed by the 
self-adjoint extension parameters $\theta$ and $\widetilde \theta$. In this 
paper, we have not attempted to take the continuum limit by sending 
$e^2 a \rightarrow 0$. It is conceivable that the full $U(1)$ gauge symmetry
would then emerge dynamically from the manifest $\Z(k)$ symmetry, and that the
lattice theory would reduce to the non-compact doubled Chern-Simons-Maxwell 
continuum theory discussed in Section 3.

While the electric field energy does not depend on the values of the 
self-adjoint extension parameters, the magnetic field energy does. The $\Z(k)$
toric code emerges from the doubled Chern-Simons-Maxwell theory in the limit of
infinite ``photon'' mass, $M = \frac{k e^2}{2 \pi} \rightarrow \infty$. In this 
limit the magnetic field energy vanishes, and thus the dependence of the energy 
on the self-adjoint extension parameters disappears. However, the wave function 
still depends on $\theta$ and $\widetilde \theta$. As a result, the toric code 
has a large variety of hidden self-adjoint extension parameters, which form two 
$U(1)$ gauge fields, one associated with the original and one associated with 
the dual lattice. Under adiabatic changes of the parameters $\theta$ and 
$\widetilde \theta$, a $U(k)$ Berry gauge field with non-trivial Berry 
curvature arises. Since the Berry field strength turned out to be gauge 
equivalent to an Abelian field strength, manipulating the corresponding Berry 
phases will certainly not allow universal topological quantum computation. 
Still, it will be interesting to investigate further whether manipulating these 
parameters can be utilized for other forms of quantum information processing.

In this work, we have encountered a large variety of different gauge fields.
First of all, we started out with non-compact Abelian gauge fields $A$ 
and $\widetilde A$, both in the continuum and on the lattice. Upon 
compactification of the gauge group, this gave rise to the compact $U(1)$
lattice gauge fields $\varphi$ and $\widetilde \varphi$. The group space of an 
original and dual cross-based link-pair $U(1)^2$ was endowed with an abstract 
vector potential $(a,\widetilde a)$ describing the
quantized ``magnetic'' flux represented by the level $k \in \Z$. The Polyakov
loops winding around the group space torus $U(1)^2$ then gave rise to two
self-adjoint extension parameters $\theta$ and $\widetilde \theta$, which play
the role of Aharonov-Bohm phases for the ``mechanical'' analog ``particle'' 
moving in the group space. While they are invisible at the classical level,
at the quantum level the parameters $\theta$ and $\widetilde \theta$ break the
manifest gauge symmetry from $U(1)$ down to $\Z(k)$. Still, the full $U(1)$
redundancy of the gauge description remains, because the self-adjoint extension
parameters themselves turned out to be non-dynamical background lattice gauge
fields associated with the original and dual lattice, respectively. An extended 
version of the toric code then emerges at low energies, which is a $\Z(k)$ 
lattice gauge theory. Finally, by considering adiabatic changes of the
self-adjoint extension parameters $\theta$ and $\widetilde \theta$, 
$U(k)$ Berry gauge fields arise, which are defined over the base space of the
lattice gauge fields $\theta$ and $\widetilde \theta$. All these different
gauge structures are intimately related to one another, and deserve further
study, in particular, in the context of quantum information processing.

We have often mentioned the ``mechanical'' analog of a ``particle'' moving in 
the cross-based group space torus $U(1)^2$, threaded by $k$ units of quantized
``magnetic'' flux. It would be most interesting to build such a system in the
laboratory, in order to manipulate quantum information. While building a torus 
seems like a most difficult task (at least to us), we have great confidence in
the ingenuity of AMO experimentalists. Perhaps one can manipulate trapped ions 
to mimic the physics of the cross-based ``mechanical'' analog. Using digital
quantum simulations with up to 100 gate quantum operations \cite{Lan11}, this 
has already been achieved for a single plaquette of the ordinary (i.e.\ not 
extended) $\Z(2)$ toric code \cite{Bar11}. If one could experimentally 
incorporate the background gauge fields $\theta$ and $\widetilde \theta$, 
putting together several crosses would provide us with a remarkable quantum 
system that embodies a large variety of Abelian dynamical or background gauge 
fields as well as an additional $U(k)$ Berry connection.

It would be interesting to extend our investigations to non-Abelian 
Chern-Si\-mons-Max\-well theories on the lattice. In particular, one may ask 
whether they also reduce to variants of the toric code with a discrete 
non-Abelian gauge group. We leave this problem for future studies.

\section*{Acknowledgments}

We like to thank D.\ Banerjee, M.\ Blau, M.\ Dalmonte, M.\ L\"uscher, E.\ Rico 
Ortega, A.\ Smilga, and P.\ Zoller for illuminating discussions. The research 
leading to these results has received funding from the Schweizerischer 
Na\-tio\-nal\-fonds and from the European Research Council under the European 
Union's Seventh Framework Programme (FP7/2007-2013)/ ERC grant agreement 339220.

\end{document}